\documentclass[aps,prd, nofootinbib,superscriptaddress]{revtex4}
\pdfoutput=1
\usepackage{amsmath, amsfonts, amsthm, amssymb, graphicx}
\usepackage{subfig}
\usepackage{epstopdf}
\usepackage{color}

\def\be{\begin{equation}}
\def\ee{\end{equation}}
\def\ba{\begin{eqnarray}}
\def\ea{\end{eqnarray}}

\def\bdm{\begin{displaymath}}
\def\edm{\end{displaymath}}

\def\bq{\begin{quote}}
\def\eq{\end{quote}}

\def\del{\partial}
\def\ltap{\ \raise.3ex\hbox{$<$\kern-.75em\lower1ex\hbox{$\sim$}}\ }
\def\gtap{\ \raise.3ex\hbox{$>$\kern-.75em\lower1ex\hbox{$\sim$}}\ }
\def\gl{\ \raise.5ex\hbox{$>$}\kern-.8em\lower.5ex\hbox{$<$}\ }
\def\roughly#1{\raise.3ex\hbox{$#1$\kern-.75em\lower1ex\hbox{$\sim$}}}

 at 10truept

\def\GB{{\hat{\cal{G}}}}

\newcommand{\FF}{{\it Fab-Four }}

\newcommand{\beq}{\begin{equation}}
\newcommand{\eeq}{\end{equation}}
\newcommand{\bea}{\begin{eqnarray}}
\newcommand{\eea}{\end{eqnarray}}
\newcommand{\beqa}{\begin{eqnarray}}
\newcommand{\eeqa}{\end{eqnarray}}

\newcommand{\h}{{\cal H}}
\newcommand{\n}{{\hat n}}
\newcommand{\order}{{\cal O}}

\begin{document}

\title{The cosmology of  the {\it  Fab-Four}}

\author{Edmund J. Copeland} 
\author{Antonio Padilla} 
\author{Paul M. Saffin} 
\affiliation{School of Physics and Astronomy, 
University of Nottingham, Nottingham NG7 2RD, UK} 
%\author{Christos Charmousis} 
%\affiliation{LPT, CNRS UMR 8627, Universit\'e Paris Sud-11, 91405 Orsay Cedex, France.}
%\affiliation{LMPT, CNRS UMR 6083, Universit\'e Fran\c{c}ois Rabelais-Tours, 37200, France} 

\date{\today}

\begin{abstract}
We have recently proposed a novel self tuning mechanism to alleviate the famous cosmological constant problem, based on 
the general scalar tensor theory proposed by Horndeski. The self-tuning model ends up consisting of four geometric terms in the action, with each term containing a free potential function of the scalar field; the four together being labeled as the {\it  Fab-Four}. In this paper we begin the important task of  deriving the cosmology associated with the {\it  Fab-Four} Lagrangian. Performing a phase plane analysis of the system we are able to obtain a number of fixed points for the system, with some remarkable new solutions emerging from the trade-off between the various potentials. As well as obtaining inflationary solutions we also find conventional radiation/matter-like solutions, but in regimes where the energy density is dominated by a cosmological constant, and where we do not have any explicit forms of radiation or matter. Stability conditions for matter solutions are obtained and we show how it is possible for there to exist an extended period of `matter domination' opening up the possibility that we can generate cosmological structures, and recover a consistent cosmology even in the presence of a large cosmological constant.
\end{abstract}

%\pacs{}
%\keywords{}

\maketitle

%%%%%%%%%%%%%%%%%%%%%%%%%%%%%%%%%%%%%%%%%%%%%%%%%%%%%%%%
\section{Introduction}
%%%%%%%%%%%%%%%%%%%%%%%%%%%%%%%%%%%%%%%%%%%%%%%%%%%%%%%%%%

Over the last decade or so, as we have struggled to explain the nature of dark energy that is believed to be responsible for the observed acceleration of the Universe, interest has turned to the possibility that rather than being caused by an unknown form of energy density, the acceleration could be a result of a modification of Einstein's theory of General Relativity. It has resulted in an explosion of papers in the field, see \cite{review} for a detailed review of the various approaches that have been adopted; one particularly interesting direction involves scalar-tensor combinations. It seems sensible to require that any theory maintains second order field equations, and the most general scalar-tensor theory satisfying that criteria was written down back in 1974 by Horndeski  \cite{horndeski:1974} (it has recently been rediscovered in \cite{general}). Such theories of modified gravity cover a wide range of models, ranging from Brans-Dicke gravity \cite{bdgravity} to the recent models \cite{covgal,galmodels} inspired by galileon theory \cite{galileon}; the latter being examples of higher order scalar tensor Lagrangians with second order field equations. Of course all of these models can be considered as special cases of Horndeski's original action. Once the action was rediscovered, it did not take long before a perturbative analysis of the background evolution equations was carried out \cite{Kobayashi:2011nu,DeFelice:2011hq}, which allows for a stability analysis to be performed on the various background solutions.

In \cite{Charmousis:2011bf} we obtained a new class of models arising out of Horndeski's theory on FLRW backgrounds. The new models gave a viable self-tuning mechanism for solving the (old) cosmological constant problem, at least at the classical level, by completely screening the spacetime curvature from the net cosmological constant.  In order to evade the famous no-go theorem of  Weinberg \cite{nogo}, the new solutions did not assume Poincar\'e invariance to hold at the level of the  solution (as Weinberg assumed), rather we allowed it to be broken in the scalar field sector. This is similar to a route adopted in \cite{bigalileon} where the scalar field is allowed to break Poincar\'e invariance on the self-tuning vacua, whilst maintaining a flat spacetime geometry. In \cite{Charmousis:2011bf} we provided a brief sketch of how the system works, showing that by demanding the self-tuning mechanism continues to work through phase transitions, so causing the vacuum energy to jump,  we get powerful restrictions on the allowed form of Horndeski's original theory. Whereas the  original model is complicated, with many arbitrary functions of both the scalar and its derivatives, we showed that by assuming matter is only minimally coupled to the metric (required to satisfy equivalence principle (EP) considerations) then once the model is passed through our self-tuning filter, it reduces in form to just four base Lagrangians each depending on an arbitrary function of the scalar only, coupled to a curvature term. We called these base Lagrangians {the {\it  Fab-Four}}: ${\cal L}_j$, ${\cal L}_p$, ${\cal L}_g$, ${\cal L}_r$, where the indices refer to {\it John, Paul, George} and {\it Ringo}. This was followed up in \cite{Charmousis:2011ea} with a detailed derivation of the conditions that lead to the four base Lagrangians just mentioned, in which we showed how they naturally lead to self-tuning solutions. Moreover in \cite{Charmousis:2011ea} we began to address the important question of the stability of the classical solutions to quantum corrections and demonstrated that at least heuristically the self-tuning solutions can be guaranteed to receive only small quantum corrections thereby not spoiling the self-tuning nature of the solutions. 

The purpose of this paper is to begin the discussion of the cosmology associated with the {\it  Fab-Four} Lagrangian. Without a sensible cosmology the model is nothing other than an interesting aside that may give some feeling as to how the cosmological constant can be addressed, but in itself does not have anything to say about our Universe. This is a non-trivial exercise.  Note that some aspects of \FF cosmology were touched upon in \cite{j+g}.

Recall from \cite{Charmousis:2011bf} that we are dealing with situations where the net cosmological constant may be large compared to any other energy density in the system. In the conventional cosmological scenario this would inevitably lead to a period of rapid acceleration, with no prospect of a radiation or matter dominated period, hence no chance for nucleosynthesis to take place or structures to form in our Universe. We will require the four potential functions in the {\it  Fab-Four} action to act together and conspire to alleviate the influence of a net large cosmological constant before the final self tuning solution is reached. After all, we do not live in that solution just yet. To attack the problem, we will rewrite the equations of motion for the scalar field and the Friedmann equation in terms of a dynamical system allowing us to look for late time attractor solutions and to determine the stability of those solutions. 

The initial conclusions are positive, we are able to find combinations of the four potentials that do indeed lead to inflation/radiation/matter dominated like periods, with the latter two entering the self-tuning regime at very late times. For matter domination like behaviour, we find that there are  solutions  that are perturbatively stable.  These solutions are remarkable. Because we are interested in the case where the cosmological constant dominates the source, we have focussed on the case where no additional sources are present. Thus, we are able to find perturbatively stable matter-like solutions that are driven by a cosmological constant.  What is happening is that the scalar field is working to screen the pressure component of the cosmological constant before its energy density.  Eventually it will also screen the energy density, but the potentials allow for an intermediate period in which $\Lambda$ essentially behaves like cold dark matter. This is the main result of this paper.

The layout is as follows: in section \ref{horndeski-action} we briefly recap the key Hamiltonian and scalar field equations of motion for the {\it  Fab-Four} system arising from the original action of Horndeski  \cite{horndeski:1974} that is minimally coupled to matter. We do not rederive them, rather direct the reader to \cite{Charmousis:2011ea} for a rigorous derivation of the Lagrangian and evolution equations.  
We begin exploring the cosmology of this self-tuning scenario  in section \ref{sec-individual}, focussing on how each member of the \FF behaves in isolation. To see how the various members behave in combination we rewrite the field equations as first order equations using a dynamical systems approach in section \ref{self-tune-cosmology}.  We switch off curvature in order to focus on the cosmological epoch prior to self-tuning, and find scaling solutions corresponding to different types of cosmology such as radiation domination, matter domination and inflation. Strictly speaking, some of these matter-like solutions are only fixed points for vanishing cosmological constant.  Even so, as we show in section \ref{sec-summary}, both analytically and through numerical simulations, they still provide an excellent approximation to the true cosmology even when there is a large non-vanishing cosmological constant.

The reader not overly concerned with the details of how we arrived at interesting classes of \FF potentials should probably skip section \ref{self-tune-cosmology} and proceed directly to section \ref{sec-summary}.  Here we summarize the main findings of section \ref{self-tune-cosmology}, as well as providing numerical simulations of solutions when spatial curvature is turned on.  For the matter-like solutions, we also consider cosmological perturbations to weed out any problems with ghost and/or gradient instabilities. Whilst some solutions are unstable, others are  perfectly well behaved. We conclude in section \ref{conc}.

%%%%%%%%%%%%%%%%%%%%%%%%%%%%%%%%%%%%%%%%%%%%%%%%%%%%%%%%%%
\section{The self tuning Lagrangian -- the {\it  Fab-Four}}\label{horndeski-action}
%%%%%%%%%%%%%%%%%%%%%%%%%%%%%%%%%%%%%%%%%%%%%%%%%%%%%%%%%%
Given that we are interested in the cosmology associated with the {\it  Fab-Four}, we will consider homogeneous and isotropic spatial geometries of the form,
\beq
\label{eq:cosmometric}
ds^2  =-dt^2+a^2(t)\left[\frac{dr^2}{1-kr^2}+r^2(d\theta^2+\sin^2\theta\;d\phi^2)\right],
\eeq
with $k$ the constant denoting the spatial curvature.
%The following useful identities then follow,
%\ba
%\nabla^\mu\nabla_\nu\phi&=&diag\left(-\ddot\phi,-H\dot\phi,-H\dot\phi,-H\dot\phi\right) \label{eq:box-phi}\\
%\label{eq:ricciTensor}
%R^\mu_{\;\;\nu}         &=&diag\left(3\frac{\ddot a}{a},\frac{\ddot a}{a}+2H^2+2\frac{k}{a^2},\frac{\ddot a}{a}+2H^2+2\frac{k}{a^2},\frac{\ddot a}{a}+2H^2+2\frac{k}{a^2}\right)\\
%\nabla^\mu\nabla_\mu\phi&=&-\ddot\phi-3H\dot\phi\\
%R                       &=&6\left(\frac{\ddot a}{a}+H^2+\frac{k}{a^2}\right)\\
%\label{eq:rho}
%\rho&\equiv& \nabla_\mu \phi \nabla^\mu \phi = -\dot\phi^2
%\ea
%%%%%%%%%%%%%%%%%%%%%%%%%%%%%%%%%%%%%%%%%%%%%%%%%%%%%%%%%%
%\section{self tuning in scalar-tensor theories} \label{self-tune}
%%%%%%%%%%%%%%%%%%%%%%%%%%%%%%%%%%%%%%%%%%%%%%%%%%%%%%%%%%
In  \cite{Charmousis:2011bf,Charmousis:2011ea} we derived the sector of Horndeski's theory \cite{horndeski:1974} that exhibits self-tuning.  What does that mean? Given our cosmological background in vacuum (\ref{eq:cosmometric}), and the expectation that the matter sector can contribute a constant vacuum energy density, we  identify with the net cosmological constant, $\rho_\Lambda=\rho^{bare}_\Lambda+\langle \rho_{m}\rangle_\textrm{vac} $. In a self tuning scenario the net cosmological constant should not have an impact on the spacetime curvature, so whatever the value of $\rho_\Lambda$,  we still want to have a portion of flat spacetime{\footnote{Given the value of the tiny observed cosmological constant, for our purposes, flat spacetime is a very good approximation.}}. What makes it applicable to cosmology is that this argument should hold when the matter sector goes through a phase-transition, changing the overall value of $\rho_\Lambda$ by a constant amount, for example GUT phase transitions, EWK phase transitions etc. For this to work, any abrupt change in the matter sector has to be completely absorbed by the scalar field leaving the geometry unchanged. Hence the scalar field tunes itself to each change in  $\rho_\Lambda$ and this has to be allowed independently of the time (or epoque) of transition. The self tuning solution is Ricci flat, which tells us that at self tuning we have 
\ba
\label{flat}
H^2&=&-\frac{k}{a^2}.
\ea
For $k=0$ we have a flat slicing of Minkowski, whilst for $k<0$ we have a Milne slicing. For $k>0$ no flat spacetime slicing is possible. With this demand of a viable self-tuning mechanism we were able to place powerful restrictions on the allowed form of Horndeski's original Lagrangian \cite{Charmousis:2011ea}. Whereas the  original model is complicated, with many arbitrary functions of both the scalar and its derivatives, we showed that  a self-tuning solution dramatically restricted the Lagrangian in form to just four base Lagrangians each depending on an arbitrary function of the scalar only, coupled to a curvature term. We called these base Lagrangians {the {\it  Fab-Four}}, given by the following
  \begin{eqnarray}
\label{eq:john}
{\cal L}_{j} &=& \sqrt{-g} V_{j}(\phi)G^{\mu\nu} \nabla_\mu\phi \nabla_\nu \phi, \\
\label{eq:paul}
{\cal L}_{p} &=&\sqrt{-g}V_{p}(\phi)   P^{\mu\nu\alpha \beta} \nabla_\mu \phi \nabla_\alpha \phi \nabla_\nu \nabla_\beta \phi, \\
\label{eq:george}
{\cal L}_{g} &=&\sqrt{-g}V_{g}(\phi) R, \\
\label{eq:ringo}
{\cal L}_{r} &=& \sqrt{-g}V_{r}(\phi) \GB,
\end{eqnarray}
where $R$ is the Ricci scalar, $G_{\mu\nu}$ is the Einstein tensor, $P_{\mu\nu\alpha \beta}$ is the double dual of the Riemann tensor \cite{mtw}, and $\GB=R^{\mu\nu \alpha \beta} R_{\mu\nu \alpha \beta}-4R^{\mu\nu}R_{\mu\nu}+R^2$ is the Gauss-Bonnet combination. Provided that $\{V_{j}, V_{p}, V_{g}\} \neq \{0,0, constant\}$ then these Lagrangian's naturally lead to the self-tuning solutions. Intriguingly, it follows that this constraint means that General Relativity is {\it not} a {\it {\it  Fab-Four}} theory, which in itself is consistent with the fact that it does not have self-tuning solutions.  By ``self-tuning", we mean that 
\begin{itemize}
\item the theory should admit a Minkowski vacuum for any value of the net cosmological constant
\item this should remain true before and after any phase transition where the cosmological constant jumps instantaneously by a finite amount.
\item the theory should permit a non-trivial cosmology
\end{itemize}
Of course, this last condition ensures that Minkowski space is not the only cosmological solution available, which is important because we know from observation that the universe had to leave Minkowski space during its evolution. Fortunately this last condition is still  allowed because the cosmological field equations are dynamical, with the Minkowski solution corresponding to a late time fixed point, meaning that once we are on a Minkowski solution, we stay there -- otherwise we evolve to it dynamically. Note that for a homogeneous scalar, self-tuning is only possible for a Milne slicing of Minkowski. Indeed, the rate at which self-tuning kicks in and the solutions evolve towards the Milne Universe is controlled by magnitude of the spatial curvature, $|k|$. This will be evident from the numerical plots shown in section \ref{sec-summary}.

We begin our analysis of the cosmology by writing down the Hamiltonian density and scalar field equations of motion arising from the {\it  Fab-Four} Lagrangian. Using the line element (\ref{eq:cosmometric}) the Lagrangians (\ref{eq:john}) - (\ref{eq:ringo})  become
\ba \label{Lag-mini-superspace}
{\cal L}_{j}&=&3a^3V_{j}(\phi) \dot\phi^2(H^2+k/a^2) \label{Lag-mini-superspace-j},\\
{\cal L}_{p}&=&-3a^3V_{p}(\phi) \dot\phi^3H(H^2+k/a^2)\label{Lag-mini-superspace-p},\\
{\cal L}_{g}&=&-6a^3V_{g}(\phi) (H^2-k/a^2)-6a^3V_{{g},\phi}(\phi) H\dot\phi \label{Lag-mini-superspace-g},\\
{\cal L}_{r}&=&-8a^3V_{{r},\phi}(\phi)\dot\phi H(H^2+3k/a^2) \label{Lag-mini-superspace-r},
\ea
where ``dot" corresponds to differentiation with respect to $t$, $H=\frac{\dot a}{a}$ is the Hubble parameter, and  $V_{{r},\phi}(\phi) \equiv \frac{dV_{r}(\phi)}{d\phi} $ etc.
Using $E^{(\phi)}_{j}=-\frac{d}{dt}\left(\frac{\del{\cal L}_{j}}{\del\dot\phi}\right)+\frac{\del {\cal L}_{j}}{\del\phi}$, ${\cal H}_{j} = \left(\frac{\del{\cal L}_{j}}{\del\dot\phi}\right)\dot\phi + \left(\frac{\del{\cal L}_{j}}{\del\dot a}\right)\dot a - {\cal L}_{j}$ etc, we find that  the  Hamiltonian density in the presence of a matter source $\rho_{m}$ is
\be \label{hamff}
{\cal H}={\cal H}_{j}+{\cal H}_{p}+{\cal H}_{g}+{\cal H}_{r}+\rho^{bare}_\Lambda=-\rho_m,
\ee
where
\ba
&&{\cal H}_{j}=3V_{j}(\phi)\dot\phi^2\left(3H^2+\frac{k}{a^2}\right), \label{hamff-j}\\
&&{\cal H}_{p}=-3V_{p}(\phi)\dot\phi^3H\left(5H^2+3\frac{k}{a^2}\right),\label{hamff-p}\\
&&{\cal H}_{g}=-6V_{g}(\phi)\left[\left(H^2+\frac{k}{a^2}\right)+H\dot\phi \frac{V_{{g},\phi}(\phi)}{V_{g}(\phi)}\right],\qquad \label{hamff-g}\\
&&{\cal H}_{r}=-24V_{{r},\phi}(\phi)\dot\phi H\left(H^2+\frac{k}{a^2}\right). \label{hamff-r}
\ea
%Recalling the requirement that the \ons Hamiltonian density should {\it not be independent} of $\dot \phi$, it follows that lugging $H^2=-k/a^2$ into (\ref{hamff}), leads immediately to the conclusion that 
%\be \label{condV1}
%\{V_{j}, V_{p}, V_{g}\}  \neq \{0, 0, constant \}
%\ee
%which rules out General Relativity as it corresponds precisely to this forbidden combination. This makes sense, because as is well known, General Relativity is {\it not} a self-tuning theory. It also rules out the possibility of a self-tuning theory supported entirely by Ringo. 
%The point is that Ringo cannot give rise to a self-tuning theory ``without a little help from his friends", John, Paul, and George. When this is the case Ringo does have a non-trivial effect on the cosmological dynamics, but does not spoil spoil self-tuning.
The scalar equation of motion is given by 
\be \label{ephiff}
E^{(\phi)}={ E}^{(\phi)}_{j}+{ E}^{(\phi)}_{p}+{ E}^{(\phi)}_{g}+{ E}^{(\phi)}_{r} =0,
\ee
where 
\ba
&&{ E}^{(\phi)}_{j}= 6{d \over dt}\left[a^3V_{j}(\phi)\dot{\phi}\Delta_2\right]  - 3a^3V_{{j},\phi}(\phi)\dot\phi^2\Delta_2,
  \label{ephiff-j}\\
&&{E}^{(\phi)}_{p}= -9{d \over dt}\left[a^3V_{p}(\phi)\dot\phi^2H\Delta_2\right]  +3a^3V_{{p},\phi}(\phi)\dot\phi^3H\Delta_2,
 \label{ephiff-p}\\
&&{ E}^{(\phi)}_{g}= -6{d \over dt}\left[a^3V_{{g},\phi}(\phi)\Delta_1\right]  +6a^3V_{{g},\phi\phi}(\phi)\dot\phi \Delta_1 ,
+6a^3V_{{g},\phi}(\phi)\Delta_1^2   \label{ephiff-g}\\
&&{ E}^{(\phi)}_{r} = -24 V_{{r},\phi}(\phi) {d \over dt}\left[a^3\left(\frac{k}{a^2}\Delta_1 +\frac{1}{3}  \Delta_3 \right) \right],   \label{ephiff-r}
\ea
and we have defined the quantity 
\be
\Delta_n=H^n-\left(\frac{\sqrt{-k}}{a}\right)^n,
\ee
which vanishes %for $n>0$ 
when we are on the self-tuning solution. As a result, it is easy to see that  $E^{(\phi)}$ also vanishes automatically during self-tuning. However, we note that the condition for self-tuning requires that the full scalar equation of motion should {\it not be independent} of $\ddot a$, and this is important as it ensures that the self-tuning solution can be evolved to dynamically, thereby allowing for a non-trivial cosmology. 
%From equation (\ref{ephiff}), we see that it means that 
%\be \label{condV2}
%\{V_{j}, V_{p}, V_{g}, V_{r} \} \neq \{0, 0, constant, constant\},
%\ee
%a possibility that has already been ruled out by the previous condition (\ref{condV1}). 

%Eqn.~(\ref{hamff}) can be rewritten in terms of the Hubble parameter as
%\be
%\mu_3 H^3 + \mu_2 H^2 + \mu_1 H + \mu_0 =0 \label{eq:hubblecubed}
%\ee
%where 
%\ba
%\mu_3 &=& -24 \dot{\phi} V_{{ringo},\phi}(\phi) - 15 \dot{\phi}^3V_{paul}(\phi) \label{eq:mu3} \\
%\mu_2 &=& 9 V_{john}(\phi)\dot{\phi}^2 - 6 V_{george}(\phi) \label{eq:mu2} \\
%\mu_1 &=& - 3\dot{\phi}(3\dot{\phi}^2V_{paul}(\phi) + 8 V_{{ringo},\phi}(\phi)) \frac{k}{a^2} - 6 \dot\phi V_{{george},\phi}(\phi)  \label{eq:mu1} \\
%\mu_0 &=& 3(V_{john}(\phi)\dot{\phi}^2 - 2 V_{george}(\phi))\frac{k}{a^2} + \left[\rho_\Lambda+\rho_\textrm{matter}\right]  \label{eq:mu0}
%\ea
What is it we would like to recover from these equations? Ideally we want  to find a cosmology consistent with observations,  that does not rely on any particular value for the cosmological constant. It should be able to accommodate an early period of inflation driven by some combination of the four potentials, followed by an extended period of radiation and matter domination during which nucleosynthesis could take place and in which structures could form. This would be followed by a late period of cosmic acceleration corresponding to the dark energy domination in which we find ourselves today. This is obviously a tall order, but as we will see, something that is not beyond the {\it Fab-Four}.  In the next section, we will briefly examine the cosmological behaviour of each member of the {\it Fab-Four} in isolation to gain some intuition as to how each term will drive cosmology. This will be followed by a much more thorough analysis in section \ref{self-tune-cosmology}: we introduce the powerful formalism of dynamical systems to rewrite the dynamics as a set of first order differential equations that we can then solve for their fixed points, allowing us to obtain a new set of cosmological solutions. 
%%%%%%%%%%%%%%%%%%%%%
\section{The cosmology of each member of the {\it Fab-Four}} \label{sec-individual}
%%%%%%%%%%%%%%%%%%%%%%
To get a feel for how each member of the \FF drives cosmology, we will briefly consider how they each behave in isolation, in the presence of a net cosmological, but no additional matter excitations.  We neglect the latter because we are ultimately interested in the case where the net cosmological constant dominates the source completely, so any matter exciation will be subleading.  As a result, we  set $\rho_m=\langle\rho_m\rangle_\textrm{vac}$ in (\ref{hamff}) and define the net cosomological constant $\rho_\Lambda=\rho^{bare}_\Lambda+\langle \rho_{m}\rangle_\textrm{vac} $. 

It is convenient to rewrite the equations of motion  (\ref{hamff}) and (\ref{ephiff}) using $N=\ln a$ as our evolution parameter, as opposed to proper time.  Using $\dot \phi=H \phi'$ and $a'=a$, (where $\phi' \equiv d\phi/dN$ etc...), we find that 
\ba
 {\cal H}_{j}+{\cal H}_{p}+{\cal H}_{g}+{\cal H}_{r} &=&-\rho_\Lambda, \label{hamT} \\
{ E}^{(\phi)}_{j}+{ E}^{(\phi)}_{p}+{ E}^{(\phi)}_{g}+{ E}^{(\phi)}_{r}  &=& 0, \label{ephiT}
\ea
where 
\ba
&&{\cal H}_{j}=3V_{j}(\phi)\phi'^2 H^2 \left(3H^2+\frac{k}{a^2}\right), \label{hamff-j1}\\
&&{\cal H}_{p}=-3V_{p}(\phi)\phi'^3H^4\left(5H^2+3\frac{k}{a^2}\right),\label{hamff-p1}\\
&&{\cal H}_{g}=-6V_{g}(\phi)\left[\left(H^2+\frac{k}{a^2}\right)+H^2  \frac{V_{{g},\phi}(\phi)\phi' }{V_{g}(\phi)}\right],\qquad \label{hamff-g1}\\
&&{\cal H}_{r}=-24V_{{r},\phi}(\phi)\phi' H^2 \left(H^2+\frac{k}{a^2}\right), \label{hamff-r1}
\ea
and
\ba
&&{ E}^{(\phi)}_{j} \dot \phi/H=\frac{1}{a^3 \Delta_2} \left[\frac{{\cal H}_j a^6 \Delta_2^2}{3 \Delta_2-2\frac{k}{a^2}} \right]'
  \label{ephiff-j1}\\
&&{E}^{(\phi)}_{p}\dot \phi/H=\frac{2}{a^{3/2} (H \Delta_2)^{1/2}} \left[\frac{{\cal H}_p a^{9/2} (H\Delta_2)^{1/2} \Delta_2 }{5\Delta_2-2\frac{k}{a^2}} \right]'
 \label{ephiff-p1}\\
&&{ E}^{(\phi)}_{g}\dot \phi/H=-3\frac{V_g'}{a} (\Delta_2 a^4)'   \label{ephiff-g1}\\
&&{ E}^{(\phi)}_{r} \dot \phi/H= \frac{{\cal H}_r}{H \Delta_2} \left[a^3\left(\frac{k}{a^2}\Delta_1 +\frac{1}{3}  \Delta_3 \right) \right]'.   \label{ephiff-r1}
\ea
Now, we are interested in the case where only one member of the \FF is switched on. We therefore take ${\cal H}_i=-\rho_\Lambda=$ constant, and solve $E^{(\phi)}_i=0$  for $i=j, p, g, r$.  We know that at late times the solutions asymptote to a curvature dominated  Milne universe $H^2 \to -\frac{k}{a^2}$ by the self-tuning mechanism \cite{Charmousis:2011bf,Charmousis:2011ea}. To examine what happens before that we  consider the opposite regime in which the curvature is subdominant; the results are listed below:
\begin{align}
V_j\textrm{ only}: &&  H^2&=\frac{M^2 }{a^6}  (1+{\cal O}(k/a^2)) && \textrm{``stiff fluid''}\\
V_p\textrm{ only}: && H^2& =\frac{M^2 }{a^6}   (1+{\cal O}(k/a^2)) && \textrm{``stiff fluid'}\\
V_g\textrm{ only}: && H^2&= \frac{M^2}{a^4}-\frac{k}{a^2} && \textrm{``radiation"}\\
V_r\textrm{ only}: &&  H^2&=\frac{M^2}{a^2} (1+{\cal O}(k/a^2)) && \textrm{``curvature"}
\end{align}
where $M$ is some mass scale that arises as a constant on integration in each case. $V_j$ and $V_p$ both behave like a stiff fluid with equation of state $w=1$, which might have been expected since the corresponding Lagrangians contain derivative terms for $\phi$, and we get a kinetically driven scalar field.  In contrast, $V_g$ behaves like radiation whilst $V_r$ behaves like curvature.  It is worth emphasizing that in each case the source is a cosmological constant with equation of state $w=-1$, and yet  the resulting cosmology behaves nothing like an inflationary de Sitter solution. The scalar field screens the pressure components of the source in a number of  different ways, depending on which member of the \FF is turned on.  This allows us to be optimistic about extracting realistic cosmological solutions from the interplay between terms in the full \FF theory. In the next section, we will show using a detailed dynamical systems analysis, that this is indeed the case

%%%%%%%%%%%%%%%%%%%%%%%%%%%%%%%%%%%%%%%%%%%%%%%%%%%%%%%%%%
\section{A dynamical systems approach to the {\it  Fab-Four} Cosmology}\label{self-tune-cosmology}
%%%%%%%%%%%%%%%%%%%%%%%%%%%%%%%%%%%%%%%%%%%%%%%%%%%%%%%%%%

Our system of equations (\ref{hamff}) and (\ref{ephiff}) are complicated second order equations and generally will not present tractable solutions. A powerful method to obtain attractor solutions is to adopt a phase plane analysis through a dynamical systems approach that allows us to both reduce the order of the differential equations but to also obtain the fixed point solutions without having to obtain the full dynamics of the system (see for example \cite{Halliwell:1986ja,Copeland:1997et}). This is a particularly powerful technique when we know what sort of fixed point solutions we are aiming for, in our case they correspond to radiation or matter domination or inflationary expansions. The reader who is simply interested in the cosmological solutions rather than their derivation may want to skip this section, and proceed directly to section \ref{sec-summary}.

We continue with $N=\ln a$ as the ``time" variable, 
and note that we will regularly come across  the following combinations
\be \label{sigma-def}
h=\frac{H'}{H},~~~~~~~\sigma=\frac{\sqrt{-k}}{Ha},
\ee
from which it follows that when $h=const$ we have
\ba
H&=&H_0a^h\\
\label{eq:aDepOn_h}
a&=&a_0(t-t_\star)^{-1/h},
\ea
with $H_0, a_0$ and $t_\star$ as constants of integration; these will be used when we examine the fixed-points. If we were to neglect the spatial curvature then matter-like expansion would correspond to $h=-3/2$ and radiation-like expansion to $h=-2$. Curvature domination corresponds to $h=-1$ and an inflationary trajectory would correspond to the region $-1<h<0$, with the solution approaching de Sitter expansion as $h \to 0$. We will not consider the case where we have exactly $h=0$, as it will correspond to a singular limit of our system. 

Another useful quantity when trying to understand particular solutions is the deceleration parameter that is given by
\ba
q&=&-\frac{a\ddot a}{\dot a^2}.
\ea
So for $a\sim t^p\sim t^{-1/h}$ we get
\ba\label{eq:q}
q&=&-\frac{p(p-1)}{p^2}=-(h+1).
\ea
The system of equations corresponding to  (\ref{hamff}) and (\ref{ephiff}) are quite complicated and so we require a few new variables to fully establish the dynamical system. We do this by introducing 
\be
x=H^\alpha\phi', \label{x-def}\ee
and 
\begin{align}
&&y_j=H^{\beta_j}V_j,  &&
y_p=H^{\beta_p}V_p,&&
y_g=H^{\beta_g}V_g, &&
y_r=H^{\beta_r}V_{r,\phi}, \label{yr-def}\\
&&\lambda_j=H^{\gamma_j}\frac{V_{j,\phi}}{V_j}, &&
\lambda_p=H^{\gamma_p}\frac{V_{p,\phi}}{V_p}, &&
\lambda_g=H^{\gamma_g}\frac{V_{g,\phi}}{V_g}, &&
\lambda_r=H^{\gamma_r}\frac{V_{r,\phi\phi}}{V_{r,\phi}, \label{lamdar-def}}\\
&& \mu_j=\frac{V_jV_{j,\phi\phi}}{(V_{j,\phi})^2},&&
\mu_p=\frac{V_pV_{p,\phi\phi}}{(V_{p,\phi})^2}, &&
\mu_g=\frac{V_gV_{g,\phi\phi}}{(V_{g,\phi})^2}, &&
\mu_r=\frac{V_{r,\phi}V_{r,\phi\phi\phi}}{(V_{r,\phi\phi})^2},\label{mur-def}
\end{align}
where $\alpha, \beta_i$ and $\gamma_i, (i=j,p,g,r)$ are constants. 

As in the previous section we specialize to the case where there is no matter excitation present, since we expect the net cosmological constant to dominate the source.  Although the source has equation of state $w=-1$, we will find that a judicious choice of {\it Fab-Four} potentials can mimic a standard cosmological evolution with {\it any} constant equation of state. In particular, we will show that   even in the absence of an explicit matter fluid, and in the presence of a large cosmological constant $\rho_\Lambda$, solutions  exist that evolve as if the universe is matter dominated i.e. $h =-3/2$ in the language described above. 

Substituting (\ref{x-def})-(\ref{mur-def}) into 
(\ref{hamff}) and (\ref{ephiff}) we demand that the resulting system of equations are autonomous, which requires all the factors of the Hubble parameter $H$ to scale out of the system. This follows from (\ref{hamff}) where the term $\rho_\Lambda$ is independent of $H$. Hence all the other terms need to be independent of $H$ as well. Demanding that, we find the following relations
\ba 
\beta_j&=&4-2\alpha, \label{eq:beta-j}\\
\beta_p&=&6-3\alpha, \label{eq:beta-p}\\
\beta_g&=&2, \label{eq:beta-g}\\
\beta_r&=&4-\alpha, \label{eq:beta-r}\\
\gamma_j=\gamma_p=\gamma_g=\gamma_r&=&-\alpha. \label{eq:gamma}
\ea
The individual Hamiltonian parts are 
\ba 
{\cal H}_j&=&3x^2y_j(3-\sigma^2), \label{eq:ham-j}\\
{\cal H}_p&=&-3x^3y_p(5-3\sigma^2),\label{eq:ham-p}\\
{\cal H}_g&=&-6y_g\left[(1-\sigma^2)+\lambda_gx\right],\label{eq:ham-g}\\
{\cal H}_r&=&-24xy_r(1-\sigma^2),\label{eq:ham-r}
\ea
and combine in (\ref{hamff}) to become 
\ba \label{eq:Hconstraint}
{\cal H}&=&{\cal H}_j+{\cal H}_p+{\cal H}_g+{\cal H}_r+\rho_\Lambda=0.
\ea
Similarly, the individual pieces of the scalar field equation of motion (\ref{ephiff}) are now 
\ba
E^{(\phi)}_j/(3a^3H^\alpha)&=&6xy_j(1-\sigma^2)+2(\alpha-1)hxy_j(1-\sigma^2)+2[xy_j(1-\sigma^2)]'-\lambda_jx^2y_j(1-\sigma^2), \label{eq:Ephi-j}\\
E^{(\phi)}_p/(3a^3H^\alpha)&=&-9x^2y_p(1-\sigma^2)-3(\alpha-1)hx^2y_p(1-\sigma^2)-3[x^2y_p(1-\sigma^2)]'+\lambda_px^3y_p(1-\sigma^2),\label{eq:Ephi-p}\\
E^{(\phi)}_g/(3a^3H^\alpha)&=&-6\lambda_gy_g(1-\sigma)+2(1-\alpha)h\lambda_gy_g(1-\sigma)-2[\lambda_gy_g(1-\sigma)]'\nonumber \\
                                &~&+2\mu_g\lambda^2_gy_gx(1-\sigma)+2\lambda_gy_g(1-\sigma)^2, \label{eq:Ephi-g}\\
E^{(\phi)}_r/(3a^3H^\alpha)&=&-24y_r(1+h)\left[-\sigma^2(1-\sigma)+\frac{1}{3}(1-\sigma^3)\right]-8y_r\left[-\sigma^2(1-\sigma)+\frac{1}{3}(1-\sigma^3)\right]', \label{eq:Ephi-r}
\ea
and give the full scalar equation of motion
\be \label{eq:Ephi}
E^{(\phi)}={ E}^{(\phi)}_{j}+{ E}^{(\phi)}_{p}+{ E}^{(\phi)}_{g}+{ E}^{(\phi)}_{r}=0.
\ee 
For completeness we can write down the individual pieces of the Euler-Lagrange equations for the scale factor, 
$E^{(a)}_{j}=\frac{\del {\cal L}_{j}}{\del a}-\frac{d}{dt}\left(\frac{\del{\cal L}_{j}}{\del\dot{a}}\right)+\frac{d^2}{dt^2}\left(\frac{\del{\cal L}_{j}}{\del\ddot{a}}\right)$, etc.  which combine to give the full scale factor equation of motion
\be\label{eq:Ea}
E^{(a)} = E^{(a)}_j+E^{(a)}_p+E^{(a)}_g+E^{(a)}_r-3a^2\rho_\Lambda=0,
\ee
\ba
E^{(a)}_j&=&3a^2\left\{-4x^2y_j+2hx^2y_j-2(x^2y_j)'+x^2y_j(1-\sigma^2)\right\}, \label{eq:Ea-j}\\
E^{(a)}_p&=&3a^2\left\{2x^3y_p(3-\sigma^2)-hx^3y_p(3-\sigma^2)+[x^3y_p(3-\sigma^2)]'\right\},  \label{eq:Ea-p}\\
E^{(a)}_g&=&3a^2\left\{4y'_g+4y_g(2-h)+2(\lambda_gxy_g)'-2h\lambda_gxy_g-2y_g(1+\sigma^2)\right\},  \label{eq:Ea-g}\\
E^{(a)}_r&=&3a^2\left\{16xy_r(1-\sigma^2)-8hxy_r(1-\sigma^2)+8[xy_r(1-\sigma^2)]'\right\}.  \label{eq:Ea-r}
\ea
Of course, although we have written down the equations for the Hamiltonian constraint, the scalar field and the scale factor, we only require two of these sets  as the third can always be obtained as a combination of the other two.

We have nearly completed our derivation of the dynamical system. All that remains is to consider the evolution of the $y,\, \lambda$ and $\mu$  coefficients in (\ref{x-def}) - (\ref{mur-def}) and $\sigma$ in (\ref{sigma-def}). These are obtained directly from their definitions. All the $\lambda$'s evolve as
\ba \label{lambda-evoln}
\lambda_i'&=&\left[-\alpha h+\lambda_i x(\mu_i-1)\right]\lambda_i,~~~~~~{i=j,p,g,r}.
\ea
The $y$'s evolve according to
\ba \label{y-evoln}
y_i'&=&\left[\beta_i h+\lambda_i x\right]y_i, ~~~~~~{i=j,p,g,r}
\ea
whereas $\sigma$ evolves according to
\ba \label{sigma-evoln}
\sigma'&=&-(1+h)\sigma.
\ea
We note that fixed point solutions for $\lambda_i$  and $y_i$ correspond to 
\ba
\alpha h &=& \lambda_i x(\mu_i-1) \label{eq:alpha-h} \\
\beta_i h &=& - \lambda_i x \label{eq:beta-h} 
\ea
Similarly from (\ref{sigma-evoln}) fixed point solutions for $\sigma$ exist for $h=-1$ and $\sigma =0$. Our complete system of equations for  investigatating  the cosmology is therefore given by (\ref{eq:Hconstraint}), (\ref{eq:Ea}), (\ref{eq:Ephi}), (\ref{lambda-evoln}), (\ref{y-evoln}) and (\ref{sigma-evoln}). We also have the definitions (\ref{sigma-def}), (\ref{x-def}) - (\ref{mur-def}) and the conditions (\ref{eq:beta-j}) - (\ref{eq:gamma}). 

In order to obtain closed form  solutions, it is convenient to assume that {\bf all the $\mu_i$ are constant}, which generically corresponds to power law potentials
\ba
V_i&=&V_{0i}\phi^{1/(1-\mu_i)},~~~~~~{i=j,p,g} \label{eq:Vjpg}, \\
V_{r,\phi}&=&V_{0r}\phi^{1/(1-\mu_r)}. \label{eq:Vringo}
\ea
The special value of $\mu_i=1$ corresponds to the case where the potentials are exponential. We can now  make some immediate progress: using (\ref{eq:alpha-h}) and  (\ref{eq:beta-h}) at the fixed point we show that 
\ba
\alpha&=&\beta_i(1-\mu_i)
\ea
By further imposing  (\ref{eq:beta-j}) to (\ref{eq:beta-r}), we see that the powers in the potentials (\ref{eq:Vjpg}) and (\ref{eq:Vringo}) are given by 
\ba
\frac{1}{1-\mu_j} &=& \frac{\beta_j}{\alpha}=\frac{4}{\alpha }-2 \\
\frac{1}{1-\mu_p} &=& \frac{\beta_p}{\alpha}=\frac{6}{\alpha }-3 \\
\frac{1}{1-\mu_g} &=& \frac{\beta_g}{\alpha}=\frac{2}{\alpha } \\
\frac{1}{1-\mu_r} &=& \frac{\beta_r}{\alpha}= \frac{4}{\alpha }-1
\ea
We can also integrate (\ref{lambda-evoln}) to give
\ba
\lambda_i&=&\frac{1}{1-\mu_i}V_{0i}^{1-\mu_i}y_i^{\mu_i-1},~~~~~~{i=j,p,g,r} \label{eq:lambda-y}
\ea

\subsection{$\sigma=0$ fixed point: vanishing spatial curvature} \label{sec:sigma=0}
%%%%%%%%%%%%%%%%%%%
We shall begin by studying the fixed point $\sigma=0$  because it corresponds to the situation where spatial curvature is sub-dominant, and so we expect the early behaviour to be closely matched by such fixed points even though $\sigma$ may not strictly vanish. 
% The  Hamiltonian (\ref{eq:Hconstraint}) becomes 
% \ba \label{eq:ham-sigma=0}
% {\cal H}=9x^2y_j-15x^3y_p-6y_g\left[1+\lambda_gx\right]-24xy_r+\rho_\Lambda=0
% \ea
% The scalar field equation of motion (\ref{eq:Ephi}) becomes
% \ba\label{eq:Ephi-sigma=0}
% E^{(\phi)}&=&6xy_j+2(\alpha-1)hxy_j+2[xy_j]'-\lambda_jx^2y_j\\\nonumber
% &~&-9x^2y_p-3(\alpha-1)hx^2y_p-3[x^2y_p]'+\lambda_px^3y_p\\\nonumber
% &~&-6\lambda_gy_g+2(1-\alpha)h\lambda_gy_g-2[\lambda_gy_g]'+2\mu_g\lambda^2_gy_gx+2\lambda_gy_g\\\nonumber
% &~&-8y_r(1+h)\\\nonumber
% &=&0,
% \ea
% and the scale factor equation of motion (\ref{eq:Ea}) is 
% \ba\label{eq:Ea-sigma=0}
% E^{(a)}&=&3a^2\left\{-4x^2y_j+2hx^2y_j-2(x^2y_j)'+x^2y_j\right.\\\nonumber
% &~&\qquad+6x^3y_p-3hx^3y_p+3[x^3y_p]'\\\nonumber
% &~&\qquad+4y'_g+4y_g(2-h)+2(\lambda_gxy_g)'-2h\lambda_gxy_g-2y_g\\\nonumber
% &~&\qquad+\left.16xy_r-8hxy_r+8[xy_r]'-\rho_\Lambda\right\}\\\nonumber
% &=&0.
% \ea
% The $\lambda$'s and $y$'s  evolve according to (\ref{lambda-evoln}) and (\ref{sigma-evoln}) with the fixed point solutions satisfying (\ref{eq:alpha-h}) and (\ref{eq:beta-h}). This in turn allows us 
Now (once again under the assumption that all the $\mu_i$ are well-defined and constant), we can make use of (\ref{eq:lambda-y}) and (\ref{eq:alpha-h}) to obtain
\ba
 y_i&=& V_{0i}[(1-\mu_i)\lambda_ix/x]^{1/(\mu_i-1)}, \label{eq:yi2}\\
    &=& V_{0i}(-\alpha h/x)^{-\beta_i/\alpha}. \label{eq:yi3}
\ea 
We therefore have $y_i$ in terms of $h$ and $x$ once we have chosen a specific $\alpha$, from which the $\mu_i$ are derived and hence the potentials determined. It allows us to replace the $y_i$ terms in the scale factor, and the scalar equations of motion at the fixed-point (where everything is constant), which leads to two equations for the variables, $x$ and $h$. The Hamiltonian is not an independent equation but could have been used instead of say the scale factor equation of motion.

Once we reach the fixed point, and using (\ref{eq:beta-j})-(\ref{eq:beta-r}), and (\ref{eq:beta-h}), the scalar field equation of motion (\ref{eq:Ephi}) becomes 
\ba
2xy_j(3+h)-3x^2y_p(3+h)-2\lambda_gy_g(2+h)-8y_r(1+h)&=&0,
\ea
which in turn becomes upon using (\ref{eq:yi3})
\ba 
-3(3+h)V_{0p}(-\alpha h)^{2-2/\alpha}x^{2/\alpha}-2V_{0j}\alpha h(3+h)-8V_{0r}(1+h)+4h(2+h)V_{og}(-\alpha h)^{-1+2/\alpha}x^{-2/\alpha}&=&0.
\ea
We could equivalently write this as
\be \label{eq:band-quadratic-form}
A(x^{2/\alpha})^2+B(x^{2/\alpha})+C=0,
\ee
where 
\ba
A&=&-3(3+h)V_{0p}(-\alpha h)^{2-2/\alpha},\\
B&=&-2V_{0j}\alpha h(3+h)-8V_{0r}(1+h),\\
C&=&4h(2+h)V_{0g}(-\alpha h)^{-1+2/\alpha},
\ea
and solve the quadratic to find $x(\alpha,h,V_{0j}/V_{0p},V_{0g}/V_{0p},V_{0r}/V_{0p})$. 
We also have the following for the scale-factor equation of motion (\ref{eq:Ea}) and Hamiltonian constraint (\ref{eq:Hconstraint}) respectively
\ba
(2h-3)x^2y_j+3(2-h)x^3y_p+2(3-2h+2h^2)y_g+8(2-h)xy_r&=&\rho_\Lambda, \\
9x^2y_j-15x^3y_p-6y_g\left[1+\lambda_gx\right]-24xy_r &=&-\rho_\Lambda. \label{Ham}
\ea
In particular the Hamiltonian (\ref{Ham})  becomes a cubic equation in $x^{2/\alpha}$
\ba\label{eq:ham-all-band}
15\alpha^3h^3V_{0p}\left(-\frac{\alpha h}{x}\right)^{-6/\alpha}+3h\alpha[3h\alpha V_{0j}+8V_{0r}]\left(-\frac{\alpha h}{x}\right)^{-4/\alpha}
-6(1-2h)V_{0g}\left(-\frac{\alpha h}{x}\right)^{-2/\alpha}+\rho_\Lambda&=&0
\ea
Now, any self tuning solution should be such that the parameters in the potential, $\alpha$, $V_{0i}, (i=j,p,r,g)$ are independent of $\rho_\Lambda$, in other words they should not be fine tuned for a particular value of $\rho_\Lambda$. This implies that the solution for $x^{2/\alpha}$ in (\ref{eq:band-quadratic-form}) will be independent of $\rho_\Lambda$ as there is no $\rho_\Lambda$ term in it. However, this would then be inconsistent when substituted into the Hamiltonian constraint (\ref{eq:ham-all-band}) where a $\rho_\Lambda$ term is present. In order to avoid this potential inconsistency we must have that each of $A$, $B$ and $C$ vanish in (\ref{eq:band-quadratic-form})  because it then does not give an equation for $x^{2/\alpha}$, and so $x$ is allowed to depend on $\rho_\Lambda$ in (\ref{eq:ham-all-band}). 

We can recover various results from (\ref{eq:band-quadratic-form}) corresponding to different scale-factor evolutions (determined by $h$ (\ref{eq:aDepOn_h})) by considering the vanishing of various terms. We list some interesting examples  in table \ref{tab-cosmologies}.
\begin{table}[h]
\caption{Examples of interesting cosmological behaviour for various fixed points with $\sigma=0$.}
\begin{tabular}{|c | c | c | c | c |  c |}
\hline
 Case & cosmological  behaviour & $V_j(\phi)$ & $V_p(\phi)$ & $V_g(\phi)$ &$V_r(\phi)$  \\
 \hline\hline
 Stiff fluid &$H^2 \propto 1/a^6$   & $c_1 \phi^{\frac{4}{\alpha}-2}$  & $c_2 \phi^{\frac{6}{\alpha}-3}$ & 0 &0 \\
 \hline 
 Radiation & $H^2 \propto 1/a^4 $  & $c_1 \phi^{\frac{4}{\alpha}-2}$ & 0 & $c_2 \phi^\frac{2}{\alpha}$& $ -\frac{\alpha^2}{8} c_1 \phi^\frac{4}{\alpha}$ \\
 \hline
 Curvature & $H^2 \propto 1/a^2$ & 0 & 0 &0 & $c_1\phi^\frac{4}{\alpha}$ \\
 \hline
 Arbitrary & $H^2 \propto a^{2h},\quad h \neq 0$ & $c_1(1+h) \phi^{\frac{4}{\alpha}-2}$ & 0 & 0 & $-\frac{\alpha^2}{16}h (3+h) c_1 \phi^\frac{4}{\alpha}$ \\
 \hline
 \end{tabular}
 \label{tab-cosmologies}
\end{table}
The first three of these are consistent with the results of the previous section. However, perhaps the most interesting solution is that last one, labelled ``Arbitrary".  This corresponds to any cosmological evolution consistent with a power law expansion $a \propto t^{-1/h}$, including a matter dominated universe ($h=-3/2$), and  inflationary expansion ($-1<h < 0$). 

We will discuss the behaviour of these solutions in more detail in section \ref{sec-summary}.  In particular we will switch the spatial curvature back on, and use  numerical  simulations to  demonstrate  how these various cosmological behaviours dominate at early times before giving in to self tuning at late times\footnote{Such late time behaviour only applies to matter and radiation, but not inflation}, and so an asymptotically Milne Universe. We will also show that the most interesting of these solutions -- the matter-like cosmology -- suffers from a rapid gradient instability at the level of cosmological perturbations.  This is unfortunate, but all is not lost. In the following section we obtain a new class of matter-like scaling solutions that exist for $\rho_\Lambda =0$.  This may seem like a strange thing to do given that the \FF was introduced to deal with a large cosmological constant, however, the analysis of the forthcoming section is simply a means to an end. As we will see in section \ref{sec-summary}, in certain cases, these solutions will not dramatically alter their behaviour when $\rho_\Lambda$ is turned on, and, more importantly, they are stable.

%%%%%%%%%%%%%%%%%%%
\subsection{$\sigma=0$ fixed point with $\rho_\Lambda=0$: vanishing curvature and cosmological constant}\label{sec:the-band-no-lambda}
%%%%%%%%%%%%%%%%%%%
Setting $\rho_\Lambda=0$ in the Hamiltonian constraint (\ref{hamff}) implies that for fixed point solutions to exist the individual terms (\ref{hamff-j})-(\ref{hamff-r}) no longer have to be time-independent. They can all scale with the Hubble parameter, for instance, as long as they scale in the same way. This then has a knock-on effect on the allowed solutions as we shall now see. In terms of the scaling parameters, the individual components of the Hamiltonian  (\ref{eq:ham-j})-(\ref{eq:ham-r}) are replaced by
\ba
{\cal H}_j&=&3x^2y_j(3-\sigma^2)H^{4-\beta_j-2\alpha} \label{eq:ham-j-n},\\
{\cal H}_p&=&-3x^3y_p(5-3\sigma^2)H^{6-3\alpha-\beta_p}\label{eq:ham-p-n},\\
{\cal H}_g&=&-6y_g\left[(1-\sigma^2)H^{2-\beta_g}+\lambda_gxH^{2-\beta_g-\alpha-\gamma_g}\right] \label{eq:ham-g-n},\\
{\cal H}_r&=&-24xy_r(1-\sigma^2)H^{4-\alpha-\beta_r}\label{eq:ham-r-n},
\ea
where the explicit $H$ dependence is included. Now if we demand that all the individual terms in the Hamiltonian scale as $H^n$ (which, for constant $h$, is equivalent to $H^n \sim a ^{nh}$) then we obtain the following constraints, replacing (\ref{eq:beta-j})-(\ref{eq:gamma})
\ba 
\beta_j&=&4-2\alpha -n, \label{eq:beta-j-n}\\
\beta_p&=&6-3\alpha -n,\label{eq:beta-p-n}\\
\beta_g&=&2 -n,\label{eq:beta-g-n}\\
\beta_r&=&4-\alpha -n,\label{eq:beta-r-n}\\
\gamma_j=\gamma_p=\gamma_g=\gamma_r&=&-\alpha. \label{eq:gamma-n}
\ea
The scalar field equations of motion  (\ref{eq:Ephi-j})-(\ref{eq:Ephi-r}) are replaced by 
\ba
E^{(\phi)}_j/(3a^3H^{\alpha+n})&=&6xy_j(1-\sigma^2)+2(\alpha+n-1)hxy_j(1-\sigma^2)+2[xy_j(1-\sigma^2)]'-\lambda_jx^2y_j(1-\sigma^2), \label{eq:Ephi-j-n}\\
E^{(\phi)}_p/(3a^3H^{\alpha+n})&=&-9x^2y_p(1-\sigma^2)-3(\alpha+n-1)hx^2y_p(1-\sigma^2)-3[x^2y_p(1-\sigma^2)]'\\\nonumber
                                &~&+\lambda_px^3y_p(1-\sigma^2), \label{eq:Ephi-p-n}\\
E^{(\phi)}_g/(3a^3H^{\alpha+n})&=&-6\lambda_gy_g(1-\sigma)-2(\alpha+n-1)h\lambda_gy_g(1-\sigma)-2[\lambda_gy_g(1-\sigma)]'\\\nonumber
                                &~&+2\mu_g\lambda^2_gy_gx(1-\sigma)+2\lambda_gy_g(1-\sigma)^2, \label{eq:Ephi-g-n}\\
E^{(\phi)}_r/(3a^3H^{\alpha+n})&=&-24y_r(1+h)\left[-\sigma^2(1-\sigma)+\frac{1}{3}(1-\sigma^3)\right]-8y_r\left[-\sigma^2(1-\sigma)+\frac{1}{3}(1-\sigma^3)\right]', \label{eq:Ephi-r-n}
\ea
the scale factor equations (\ref{eq:Ea-j})-(\ref{eq:Ea-r}) become 
\ba
E^{(a)}_j/(3a^2H^n)&=&\left\{-4x^2y_j+2(1-n)hx^2y_j-2(x^2y_j)'+x^2y_j(1-\sigma^2)\right\}, \label{eq:Ea-j-n}\\
E^{(a)}_p/(3a^2H^n)&=&\left\{2x^3y_p(3-\sigma^2)-(1-n)hx^3y_p(3-\sigma^2)+[x^3y_p(3-\sigma^2)]'\right\},  \label{eq:Ea-p-n}\\
E^{(a)}_g/(3a^2H^n)&=&\left\{4y'_g+4y_g(2-(1-n)h)+2(\lambda_gxy_g)'-2(1-n)h\lambda_gxy_g-2y_g(1+\sigma^2)\right\},  \label{eq:Ea-g-n}\\
E^{(a)}_r/(3a^2H^n)&=&\left\{16xy_r(1-\sigma^2)-8(1-n)hxy_r(1-\sigma^2)+8[xy_r(1-\sigma^2)]'\right\},  \label{eq:Ea-r-n}
\ea
and, for completeness, the Hamiltonian constraint remains as before (recall we have ensured the time dependence in $H$ factors out) 
\be \label{eq:ham-n}
3x^2y_j(3-\sigma^2) -3 x^3y_p (5-3\sigma^2) -6y_g\left[(1-\sigma^2)+\lambda_gx\right] -24xy_r(1-\sigma^2) =0.
\ee

The results obtained previously for the $\lambda$'s, $y$'s and $\sigma$ (\ref{lambda-evoln})-(\ref{eq:beta-h}) still apply, although of course with the new values for the $\beta$ coefficients. 
%\ba
%\lambda'&=&\left[-\alpha h+\lambda x(\mu-1)\right]\lambda
%\ea
%with the $y$s evolving according to
%\ba
%y'&=&\left[\beta h+\lambda x\right]y
%\ea
%along with
%\ba
%\sigma'&=&-(1+h)\sigma
%\ea
%just as before. 
We now look for $\sigma=0$ fixed point solutions in order to understand the early-time behaviour.  The Hamiltonian constraint (\ref{eq:ham-n}) becomes 
\ba \label{eq:ham-n-sigma=0}
&~&9x^2y_j
-15x^3y_p
-6y_g\left[1+\lambda_gx\right],
-24xy_r=0,
\ea
and the scalar equation of motion (\ref{eq:Ephi-j-n})-(\ref{eq:Ephi-r-n}) becomes
\ba \label{eq:Ephi-n-sigma=0}
&~&6xy_j+2(\alpha+n-1)hxy_j-\lambda_jx^2y_j -9x^2y_p-3(\alpha+n-1)hx^2y_p \nonumber \\
&~&+\lambda_px^3y_p-4\lambda_gy_g-2(\alpha+n-1)h\lambda_gy_g
                                +2\mu_g\lambda^2_gy_gx -8y_r(1+h) =0.
\ea
We can follow the route taken in the previous subsection and  use the fact that for all four potentials ($i=j,p,g,r$) (\ref{eq:alpha-h}) and (\ref{eq:beta-h}) still hold, giving
\ba
%\lambda_i x&=&-\beta_i h\\
%\lambda_i x&=&\frac{\alpha h}{\mu_i-1}\\
\Rightarrow\alpha/\beta_i&=&1-\mu_i,
\ea
which leaves (\ref{eq:Ephi-n-sigma=0}) as 
\ba \label{eq:Ephi-n-sigma=0-1}
&~&[6+(n+2)h)]xy_j
-[9+(3+2n)h]x^2y_p
-2\lambda_gy_g[2+h]-8y_r(1+h)=0.
\ea
Now, if we take $\mu=const$ then we can use (\ref{eq:yi3}) to rewrite the Hamiltonian (\ref{eq:ham-n-sigma=0}) as 
%then we know $V=V_0\phi^{1/(1-\mu)}$\footnote{In the case of $V_r$ we actually have $V_{r,\phi}=V_{0r}\phi^{1/(1-\mu_r)}$ giving $V_r=\frac{1-\mu_r}{2-\mu_r}V_{0r}\phi^{(2-\mu_r)/(1-\mu_r)}$}, in which case we find, as before,
%\ba
%y^{\mu-1}&=&(1-\mu)(V_0)^{\mu-1}\lambda\\
%\Rightarrow y&=&V_0\left[(1-\mu)\lambda\right[^{1/(\mu-1)}=V_0\left[(1-\mu)\lambda x/x\right]^{1/(\mu-1)}
%      =V_0\left[-(1-\mu)\beta h/x\right]^{1/(\mu-1)}\\\nonumber
%   &=&V_0\left(-\alpha h/x\right)^{1/(\mu-1)}
%\ea
%so
%\ba
%y&=&V_0(-\alpha h/x)^{-\beta/\alpha}
%\ea
%then the Hamiltonian becomes
%\ba
%&~&9\alpha^2h^2V_{0j}\left(-\frac{\alpha h}{x}\right)^{(n-4)/\alpha}
%+15V_{0p}\alpha^3h^3\left(-\frac{\alpha h}{x}\right)^{(n-6)/\alpha}
%-6[1+(n-2)h]V_{0g}\left(-\frac{\alpha h}{x}\right)^{(n-2)/\alpha}\\\nonumber
%&~&+24V_{0r}\alpha h\left(-\frac{\alpha h}{x}\right)^{(n-4)/\alpha}=0
%\ea
%which may be written as
\ba\label{eq:ham-n-sigma=0-1}
-6[1+(n-2)h]V_{0g}\left(-\frac{\alpha h}{x}\right)^{4/\alpha}
+[9\alpha^2h^2V_{0j}
+24\alpha hV_{0r}]\left(-\frac{\alpha h}{x}\right)^{2/\alpha}
+15\alpha^3h^3V_{0p}=0.
\ea
Similarly the scalar equation of motion (\ref{eq:Ephi-n-sigma=0-1}) may be written as
%\ba 
%&~&[6+(n+2)h)]\alpha^2 h^2V_{0j}\left(-\frac{\alpha h}{x}\right)^{(n-4)/\alpha}
%+[9+(3+2n)h]\alpha^3 h^3V_{0p}\left(-\frac{\alpha h}{x}\right)^{(n-6)/\alpha}\\\nonumber
%&~&+2(2+h)(2-n)hV_{0g}\left(-\frac{\alpha h}{x}\right)^{(n-2)/\alpha}
%+8(1+h)\alpha hV_{0r}\left(-\frac{\alpha h}{x}\right)^{(n-4)/\alpha}=0
%\ea
%giving
\ba \label{eq:Ephi-n-sigma=0-2}
&~&2(2+h)(2-n)hV_{0g}\left(-\frac{\alpha h}{x}\right)^{4/\alpha}
+\left\{[6+(n+2)h)]\alpha^2 h^2V_{0j}+8(1+h)\alpha hV_{0r}\right\}\left(-\frac{\alpha h}{x}\right)^{2/\alpha}\\\nonumber
&~&+[9+(3+2n)h]\alpha^3 h^3V_{0p}=0.
\ea
If we now define the variables
\ba
\tilde V_{0j}&=&\alpha^2 h^2V_{0j},\;\tilde V_{0p}=\alpha^3 h^3V_{0p},\;\tilde V_{0g}=V_{0g},\;\tilde V_{0r}=\alpha hV_{0r},\\
\tilde X&=&(-\alpha h/x)^{2/\alpha},
\ea
then the Hamiltonian equation (\ref{eq:ham-n-sigma=0-1}) and the scalar equation of motion (\ref{eq:Ephi-n-sigma=0-2}) become 
\ba 
-6[1+(n-2)h]\tilde V_{0g}\tilde X^2
+[9\tilde V_{0j}
+24\tilde V_{0r}]\tilde X
+15\tilde V_{0p}&=&0, \label{eq:ham-n-sigma=0-2}\\
2(2+h)(2-n)h\tilde V_{0g}\tilde X^2
+\left\{[6+(n+2)h]\tilde V_{0j}+8(1+h)\tilde V_{0r}\right\}\tilde X
+[9+(3+2n)h]\tilde V_{0p}&=&0. \label{eq:Ephi-n-sigma=0-3}
\ea
Our aim now is to find solutions of these equations. On the surface it would seem that we can just solve  for $\tilde X$ and $n$, and pick   $h$ and the $\tilde V_{0\#}$ at will. In particular, for a matter dominated epoch we would pick $h=-3/2$.  However, it is worth pausing for a moment to think about our ultimate goal. We would like any such solution to be robust against switching $\rho_\Lambda$ back on.  To see how this might be possible,  it proves useful to introduce
\ba
\hat n&=&hn,
\ea
since then, for constant $h$, we have ${\cal H} \sim H^n\sim a^{\hat n}$. So, if we can find solutions for positive $\hat n$, the correction to the Hamiltonian from switching $\rho_\Lambda$ back on, $\rho_\Lambda a^{-\hat n}$, may be expected to become less significant with expansion, meaning that the $\rho_\Lambda=0$ solution was indeed worth finding.  At this point it also proves convenient to fix the freedom we have in re-defining the scalar field, by choosing the parameter $\alpha$ such that 
\ba
\alpha h=-1.
\ea
We may then determine the powers of $\phi$ that appear in the scalar potentials (\ref{eq:Vjpg}), (\ref{eq:Vringo})
\ba
\frac{1}{1-\mu_j}&=&\beta_j/\alpha=\frac{4-n}{\alpha}-2=\hat n-(4h+2),\\
\frac{1}{1-\mu_p}&=&\beta_p/\alpha=\frac{6-n}{\alpha}-3=\hat n-(6h+3),\\
\frac{1}{1-\mu_g}&=&\beta_g/\alpha=\frac{2-n}{\alpha}=\hat n-2h,\\
\frac{1}{1-\mu_r}&=&\beta_r/\alpha=\frac{4-n}{\alpha}-1=\hat n-(4h+1).
\ea
We shall now focus on reproducing a matter epoch, so we  plug $h=-3/2$ and $\alpha=2/3$ into (\ref{eq:ham-n-sigma=0-2}) and (\ref{eq:Ephi-n-sigma=0-3}). This gives
  
  \ba 
-6(\hat n+4) V_{0g}\tilde X^2
+[9 V_{0j}
-24 V_{0r}]\tilde X
-15 V_{0p}&=&0 \label{eq:ham-n-sigma=0-2T}\\
-(\hat n+3) V_{0g}\tilde X^2
+\left[(\hat n +3) V_{0j}+4 V_{0r}\right]\tilde X
-\left(\frac{9}{2}+2\hat n\right)V_{0p}&=&0 \label{eq:ham-n-sigma=0-2T1}
\ea
We want these two equations to have a common root. Although we will miss the full spectrum of possibilities,  it is convenient and cleaner to impose the stronger constraint that the two equations have {\it two} common roots. This requires that the following equations hold for some constant $\chi$
\ba
6\chi (\hat n+4) V_{0g} &=& (\hat n+3) V_{0g},  \label{vg}\\
\chi [9 V_{0j}
-24 V_{0r}] &=& (\hat n +3) V_{0j}+4 V_{0r}, \label{vjr} \\
15 \chi V_{0p} &=& \left(\frac{9}{2}+2\hat n\right)V_{0p}. \label{vp}
\ea
Now if $V_{0p}, V_{0g} \neq 0$ then it follows from (\ref{vg}) and (\ref{vp}) that $\hat n=-\frac{3}{2}, -\frac{7}{2}$. However, as explained above, we are only interested in $\hat n>0$, so we assume that one of  $V_{0p}, V_{0g}$ vanishes. The two possibilities are presented in table \ref{tab-mattercos}.
 \begin{table}[h]
\caption{Examples of matter-like cosmological fixed points  with $\sigma=0$ and $\rho_\Lambda=0$.}
\begin{tabular}{|c | c | c | c | c |  c |}
\hline
 Case & cosmological  behaviour & $V_j(\phi)$ & $V_p(\phi)$ & $V_g(\phi)$ &$V_r(\phi)$  \\
 \hline\hline
Matter I &$H^2 \propto 1/a^3$   & $c_1 \phi^{\hat n+4}$  & $c_2 \phi^{\hat n+6}$ &  0  &  $\frac{2\hat n-3}{16(2\hat n +7)(\hat n+6)} c_1\phi^{\hat n+6}$  \\
 \hline 
 Matter II &$H^2 \propto 1/a^3$   & $c_1 \phi^{\hat n+4}$  & 0 &  $c_2 \phi^{\hat n+3}$  &  $-\frac{(\hat n+3)(2\hat n+5)}{8 (2\hat n+7)(\hat n+6)} c_1 \phi^{\hat n+6}$  \\
 \hline 
 \end{tabular}
 \label{tab-mattercos}
\end{table}
In each case, either Paul or George  is switched on/off, and the interplay of John and Ringo gives rise to a cosmology that mimics a matter dominated epoch. This is true for any value of $\hat n$, and in particular, for $\hat n>0$. We will see in the next section that these solutions are indeed robust against switching on $\rho_\Lambda$, although, again, the cosmological perturbations are unstable.

Another class of potentials can be found by  imposing the weaker constraint  that equations (\ref{eq:ham-n-sigma=0-2T}) and (\ref{eq:ham-n-sigma=0-2T1}) have just one common root. A particular example of this  is given by
\be
V_j (\phi)=-45\sqrt{2} \phi^{5}, \qquad  V_p(\phi)=-\frac{75067}{225}\phi^7, \qquad V_g(\phi)=-\phi^4, \qquad V_r(\phi)=\frac{143}{168}\sqrt{2}\phi^7
\ee
Remarkably, this leads to matter dominated  solutions that  will turn out to be stable against cosmological perturbations.

\section{Summary of cosmological solutions} \label{sec-summary}
The goal of this paper was to establish whether or not the \FF could, in principle, accommodate a consistent cosmological history. In particular, is the \FF consistent with an early period of inflation, followed by a radiation and matter dominated epoch, during which nucleosythesis takes place and structures begin to form? At late times, we want another period of inflation, before self-tuning  kicks in and the Universe enters a late time Milne solution, whatever the value of the cosmological constant. Our dynamical systems analysis, allied with the numerical solutions to be presented shortly,  demonstrates that each desired epoch can be individually realised with a judicious choice of potentials.  It is not too difficult to imagine that one could, in principle, combine the various choices in such a way that one particular choice dominates the dynamics at one particular epoch, thereby reproducing the  desired cosmic history in its entirety.  Recall  that the cosmological constant is always assumed to be large, dominating over any  other sources. Including extra sources explicitly is straightforward as we know the scaling behaviour of both radiation and matter with scale factor evolution.

%
%n the following figures we explicitly show the evolution of a number of cosmologies including the curvature term. We see how the early evolution, where the curvature is negligible, evolves according to our analysis of the $\sigma=0$ fixed-point solution, before finally self-tuning to the required late-time solution $H^2a^2 =-k$. \comment{[I dont have the files for this --Ed, I'll get them soretd later -- P]}. The key thing we have demonstrated here is the possibility of obtaining a wide class of conventional cosmologies from the {\it  Fab-Four} potentials; this is promising. Recall that we have demonstrated the existence of solutions that mimic inflation, radiation domination and matter domination simply due to the presence of the four potentials -- we have no explicit matter or radiation present. So in spite of allowing for a large $\rho_\Lambda$ contribution we can still have cosmological evolution mimicking say matter. 

Having obtained desirable solutions, we should also demonstrate that they are stable to perturbations, or at least stable enough that they will survive for the e-foldings necessary for structures to form. Scalar-tensor theories can be plagued with ghost and gradient instabilities (see \cite{review} for a review of the subject), which is one of the main reasons it is proving so challenging to develop successful modified theories of gravity.  This involves lengthy calculations in each case, and for that reason we restrict attention, for now, to the epoch of greatest interest -- matter domination. We will see that the matter epoch is perturbatively stable for the latter case. 

We will now summarise how each desired epoch can be reproduced within the {\it Fab-Four}, reinforcing our analytic calculations with numerical simulations of the full system. In each case we will plot the evolution of the deceleration parameter, $q=-a \ddot a/\dot a^2$, which is expected to give $q<0$ (inflation), $q=1$ (radiation), $q=0.5$ (matter) and $q=0$ (self-tuning). As promised, the matter epoch will also include a discussion of stability.

\subsection{The inflationary epoch}
\begin{figure}[h] 
  \centering
  \includegraphics[width=3in]{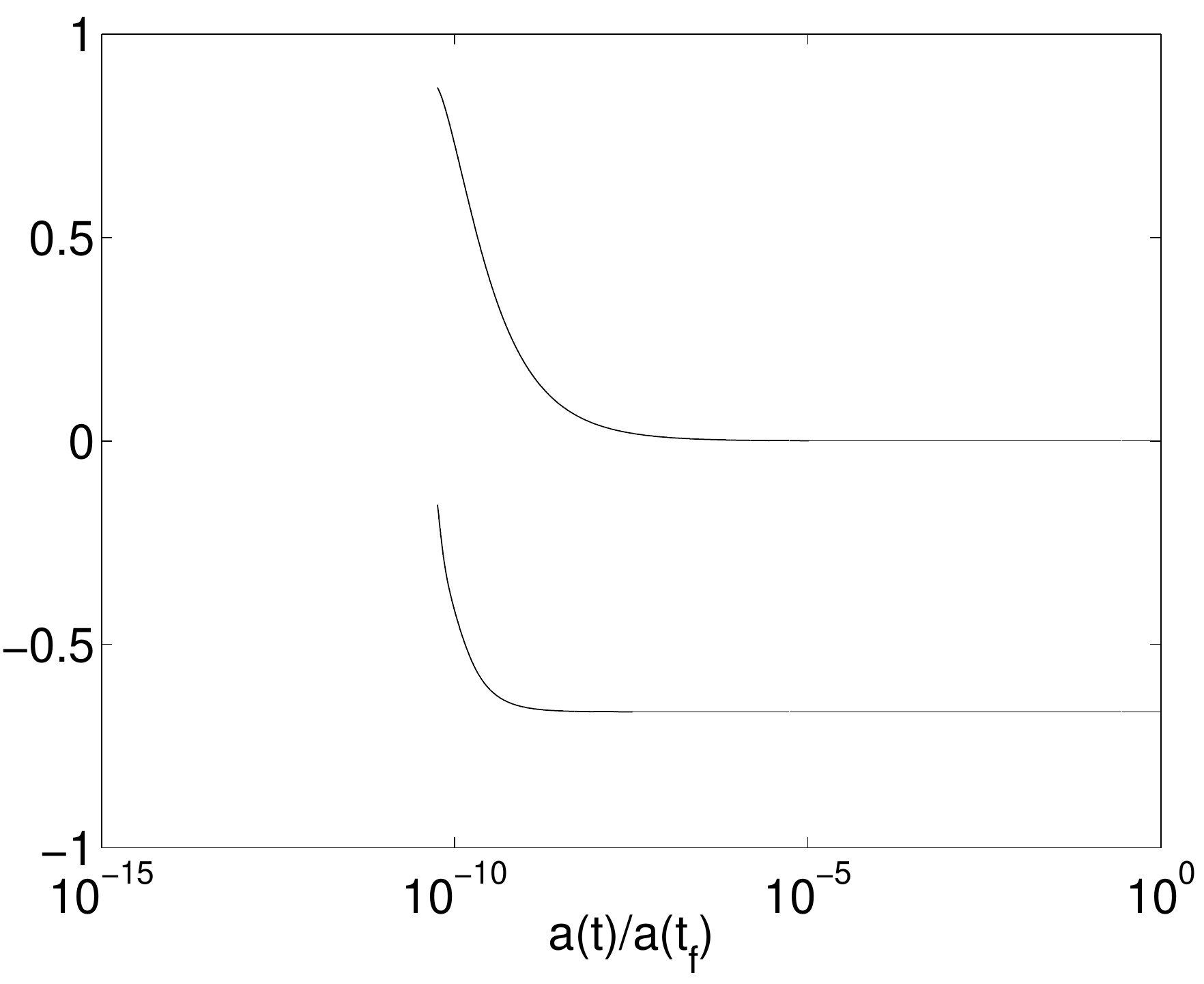} 
  \caption{Plot of the deceleration parameter, $q$, (lower curve) and $100\sigma$ (upper curve) for the inflationary case for $p=3$. The parameters used were $k=-10$, $\rho_\Lambda=1000$, $\phi_{initial}=0.1$, $c_1=1.5$}
    \label{fig-infln}
\end{figure}

It is natural ask how any self-tuning scenario can accommodate inflation, be it early or late.  The point is that self-tuning can be a late time attractor, and that inflation can happen beforehand. Clearly this would be perfectly consistent with what we observe. In section \ref{sec:sigma=0} we saw how power law inflation, $a \sim t^p$, with $p>1$ could be achieved for arbitrary $\rho_\Lambda$ with the choice:
\be
V_j (\phi)=c_1\left(\frac{p-1}{p}\right) \phi^{\frac{4}{p}-2}, \qquad  V_p(\phi)=V_g(\phi)=0, \qquad V_r(\phi)=\frac{3p-1}{16} c_1 \phi^\frac{4}{p},
\ee
where we have used the fact that $p=-1/h$, and have fixed the freedom to redefine the scalar field by setting $\alpha h=-1$. For vanishing spatial curvature, the field equations (\ref{hamT}) and (\ref{ephiT})  can be straightforwardly solved to give
\be
\phi=\nu a,  \qquad H^2=\sqrt{\frac{p \rho_\Lambda}{3 c_1 (3p+1)}} (\nu a)^{-2/p},
\ee
where $\nu$ is an arbitrary constant. The inflationary case actually presents a qualitatively different behaviour to the matter and radiation solution, as could have been guessed from (\ref{sigma-evoln}). In (\ref{sigma-evoln}) we see that one may expect $\sigma=0$ to be a repeller fixed-point for $1+h<0$, such as matter and radiation, but an attractor for $1+h>0$. Although this is rather naive, as there are many more variables to consider, this is indeed what is observed. Switching on spatial curvature and performing the full evolution drives the system to the $\sigma=0$ inflationary solution, as seen in Fig. \ref{fig-infln}. 
Of course, in a more complicated scenario in which potentials are ``sewn together" so that inflation gives way to radiation one expects self-tuning to dominate at very late times.
\subsection{The radiation epoch}
In section \ref{sec:sigma=0},  we found  that for arbitrary $\rho_\Lambda$ the following potentials will mimic a radiation dominated Universe,
\be
V_j (\phi)=c_1\phi^{6}, \qquad V_p(\phi)=0, \qquad  V_g(\phi)=c_2 \phi^4, \qquad  V_r(\phi)=-\frac{c_1}{32} \phi^8,
\ee
with a vanishing spatial curvature solution of
\be
\phi=\nu a,  \qquad H^2=\frac{c_2}{c_1}  \left[ 1\pm \sqrt{1-\frac{\rho_\Lambda c_1}{15  c_2}} \right] (\nu a)^{-4},
\ee
where $\nu$ is an arbitrary constant. Note that as $c_1 \to 0$ the negative root above has a well defined limit. This corresponds to the case where only George is turned on. We could easily have guessed that this would mimic radiation since the scalar equation of motion imposes the constraint $R=0$. This is equivalent to saying that the trace of the effective energy momentum tensor vanishes. 

\begin{figure}[h] 
  \centering
  \includegraphics[width=3in]{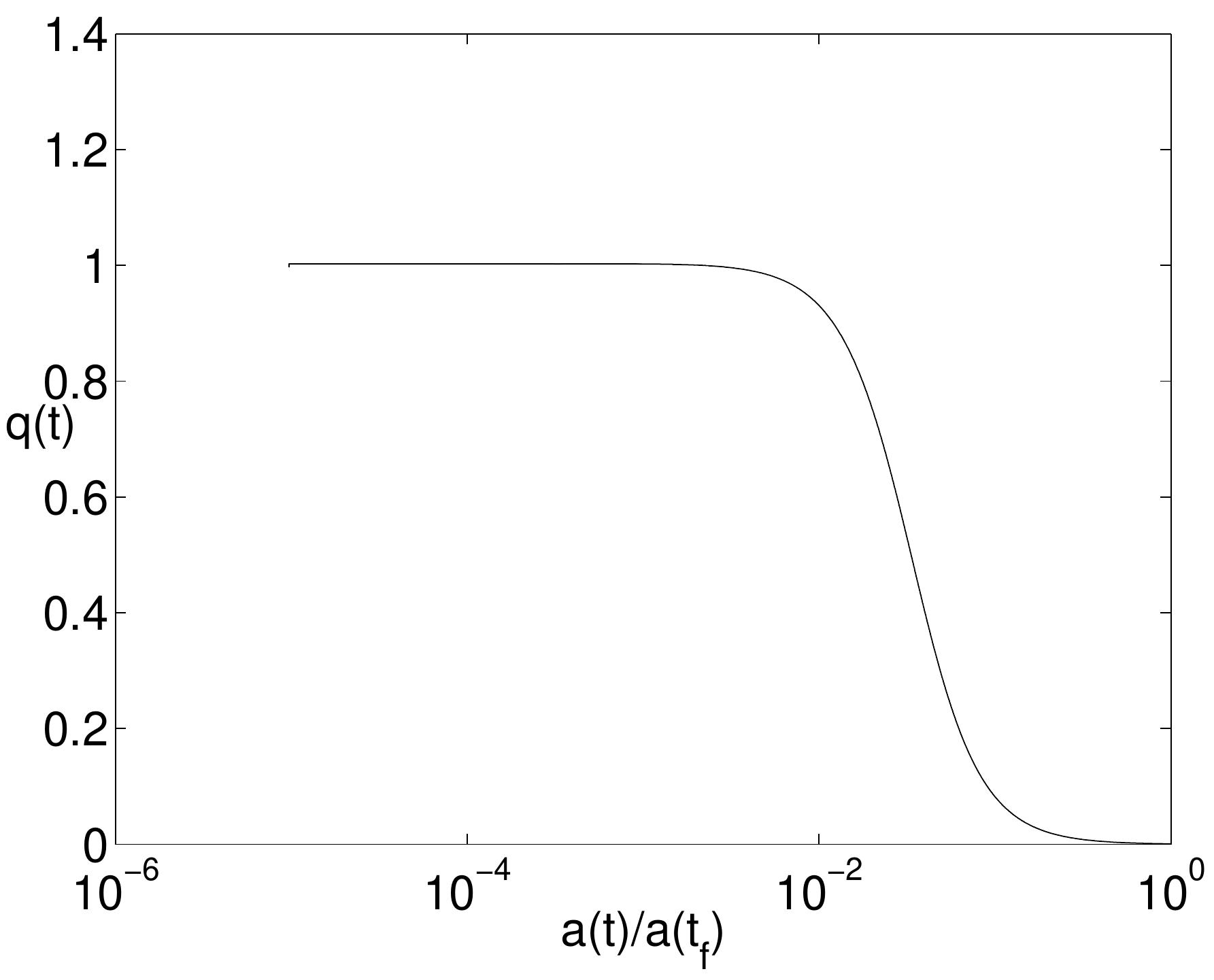} 
  \caption{Plot of the deceleration parameter for an initially radiation-like solution, with $k=-0.1$, $\rho_\Lambda=1000$, $\phi_{initial}=0.1$, $c_1=-0.5$.}
    \label{fig-rad}
\end{figure}

When we reintroduce the spatial curvature, self-tuning kicks in. This is explicitly  demonstrated by the numerical solutions presented in figure \ref{fig-rad}.

\subsection{The matter epoch}
We finally turn to the matter epoch. We have identified four classes of potential that can mimic a matter dominated universe. Let us consider the first two of these, dubbed ``Matter I" and ``Matter II", and identified in section \ref{sec:the-band-no-lambda} for vanishing curvature and cosmological constant. The ``Matter I" potentials have vanishing $V_g$, and  are given by 
\be
V_j(\phi)=c_1 \phi^{\hat n+4}, \qquad V_p(\phi)=c_2 \phi^{\hat n+6}, \qquad  V_g(\phi)= 0, \qquad V_r(\phi)=\frac{2\hat n-3}{16(2\hat n +7)(\hat n+6)} c_1\phi^{\hat n+6}\ee
For vanishing spatial curvature and cosmological constant, this admits a solution,
\be \label{sol1}
\phi=\nu a,  \qquad H^2=\left[\frac{2\hat n+9}{2(2\hat n+7)}\right] \frac{c_1}{c_2} (\nu a)^{-3} 
\ee
where $\nu$ is an arbitrary constant.  The ``Matter II" potentials have vanishing $V_p$, and  are given by 
\be
V_j(\phi)= c_1 \phi^{\hat n+4}, \qquad V_p(\phi)=  0, \qquad  V_g(\phi)= c_2 \phi^{\hat n+3}, \qquad V_r(\phi)=-\frac{(\hat n+3)(2\hat n+5)}{8 (2\hat n+7)(\hat n+6)} c_1 \phi^{\hat n+6}.
\ee
This time, for vanishing spatial curvature and cosmological constant, we have the solution,
\be  \label{sol2}
\phi=\nu a,  \qquad H^2=\left[\frac{2(2\hat n+7)}{2\hat n+9}\right] \frac{c_2}{c_1} (\nu a)^{-3},
\ee
where $\nu$ is an arbitrary constant. Note that this solution is not well defined as $c_1 \to 0$, when we expect to recover a radiation-like Universe.

\begin{figure}[h] 
  \centering
  \includegraphics[width=3in]{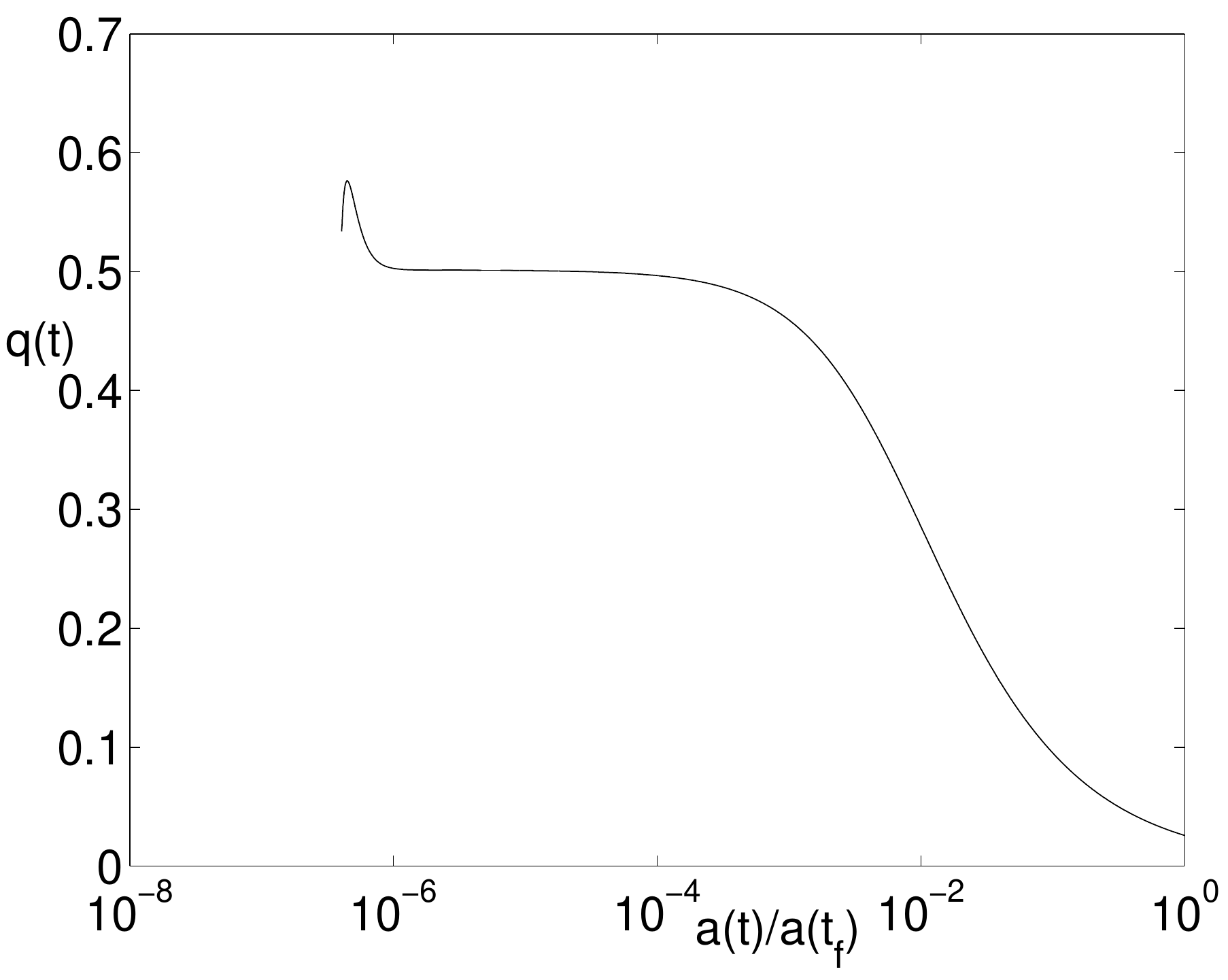} 
  \caption{Plot of the deceleration parameter for the Matter I solution, with $k=-10^{-4}$, $\phi_{initial}=3.0$, $\rho_\Lambda=1$, $\hat n=6$, $c_1=1$, $c_2=10$.}
    \label{fig-matter}
\end{figure}

\begin{figure}[h] 
  \centering
  \includegraphics[width=3in]{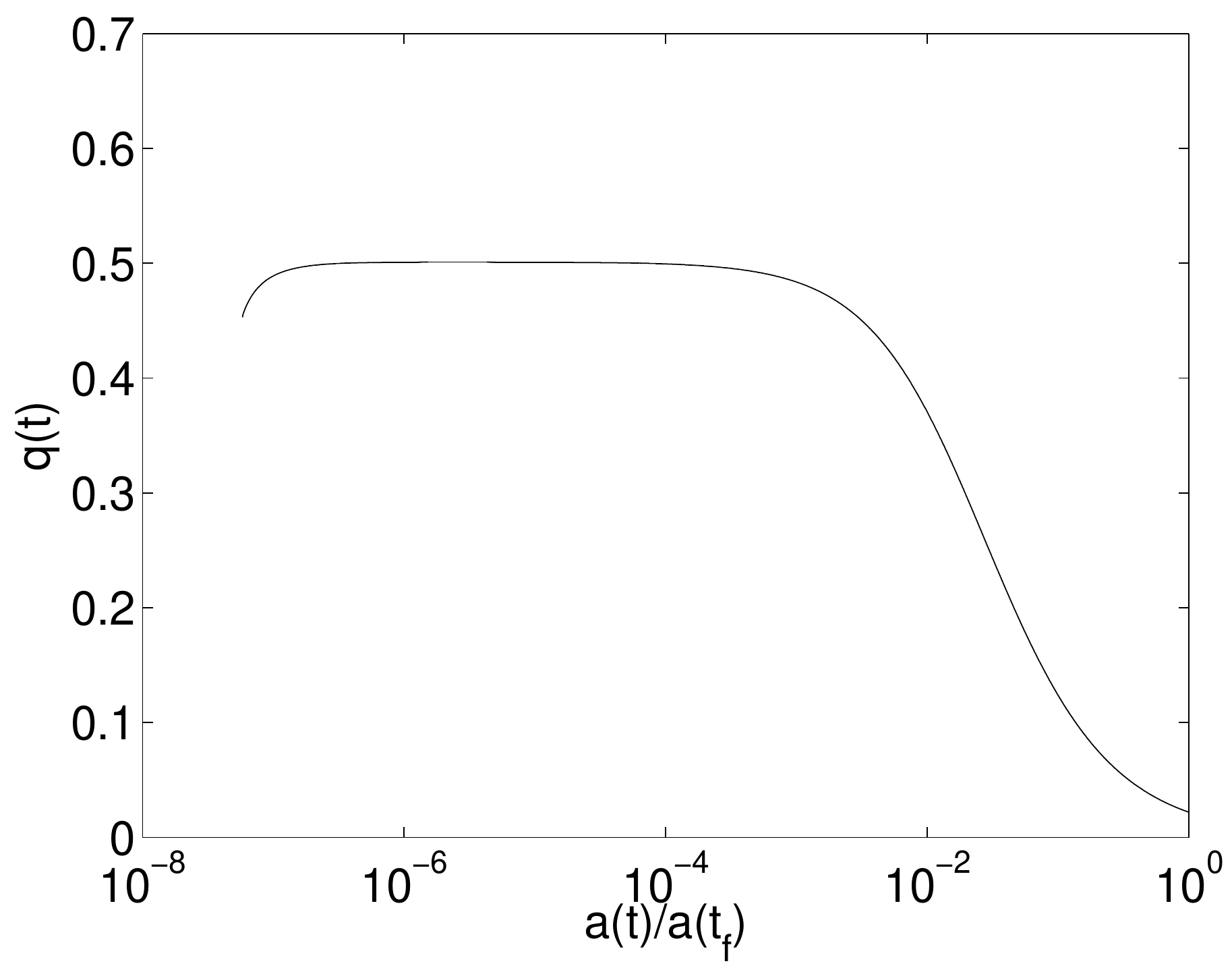} 
  \caption{Plot of the deceleration parameter for the Matter II solution, with $k=-0.1$, $\phi_{initial}=0.1$, $\rho_\Lambda=1$, $\hat n=2$, $c_1=1$, $c_2=8$.}
    \label{fig-matterii}
\end{figure}

Of course, these solutions are only valid when $\rho_\Lambda=0$, which goes against the spirit of the \FF and self-tuning.  What happens when we switch $\rho_\Lambda$ back on? To get an idea, let us perturb about these solutions and compute the Hamiltonian (\ref{hamT})  and scalar equation of motion (\ref{ephiT}) to leading order. To this end, we set
\be
\phi=\nu a +\delta \phi(N), \qquad H^2=\frac{m^2}{a^3}+ \delta H^2(N),
\ee
with 
\be
m^2=\begin{cases} \left[\frac{2\hat n+9}{2(2\hat n+7)}\right] \frac{c_1}{c_2} \nu^{-3} & \textrm{Matter I} \\
\left[\frac{2(2\hat n+7)}{2\hat n+9}\right] \frac{c_2}{c_1} \nu^{-3} & \textrm{Matter II} \end{cases}
\ee
and expand (\ref{hamT}) and  (\ref{ephiT}) to leading order in $\delta$. The result is
\ba
\delta \h_j+\delta \h_p+\delta \h_g+\delta \h_r &=&-\rho_\Lambda.  \label{hampert}\\
\delta \tilde E_j^{(\phi)}+\delta \tilde E_p^{(\phi)}+\delta \tilde E_g^{(\phi)}+\delta \tilde E_r^{(\phi)} &=&0. \label{ephipert}
\ea
where
\ba
\delta \h_j  &=& \h_j \left[ 2 \frac{\delta H^2}{H^2} +(\n+6) \frac{\delta \phi}{\phi}+2 \left(\frac{\delta \phi}{\phi}\right)' \right].\\
\delta \h_p &=& \h_p \left[ 3 \frac{\delta H^2}{H^2} +(\n+9) \frac{\delta \phi}{\phi}+3\left( \frac{\delta \phi}{\phi}\right)' \right].\\
\delta \h_g &=& \h_g \left[  \frac{\delta H^2}{H^2} +(\n+3) \frac{\delta \phi}{\phi}+\left(\frac{\n+3}{\n+4}\right) \left( \frac{\delta \phi}{\phi}\right)' \right].\\
\delta \h_r &=& \h_r \left[ 2 \frac{\delta H^2}{H^2} +(\n+6) \frac{\delta \phi}{\phi}+ \left(\frac{\delta \phi}{\phi} \right)'\right],
\ea  
and
\ba\nonumber
\delta \tilde E_j^{(\phi)}=\delta \left( E_j^{(\phi)} \dot \phi /Ha^3\right) &=& \frac{\h_j}{3} \left[ 3 \left( \frac{\delta H^2}{H^2}  \right)'+2(\n+3) \frac{\delta H^2}{H^2}+2 \left( \frac{\delta \phi}{\phi}\right)''+3(\n+4)\left( \frac{\delta \phi}{\phi}\right)'\right.\\
   &~&\left.\qquad+(\n +3)(\n+6) \frac{\delta \phi}{\phi} \right], \\\nonumber
\delta \tilde E_p^{(\phi)}=\delta \left( E_p^{(\phi)} \dot \phi /Ha^3\right) &=& \frac{2\h_p}{5} \left[ \frac{15}{4} \left( \frac{\delta H^2}{H^2}  \right)'+3\left(\n+\frac{9}{4}\right) \frac{\delta H^2}{H^2} +3  \left(\frac{\delta \phi}{\phi}\right)''+\left(4 \n+\frac{63}{4}\right)\left( \frac{\delta \phi}{\phi}\right)'\right.\\
&~&\left.\qquad+\left(\n +\frac{9}{4}\right)\left(\n+9\right) \frac{\delta \phi}{\phi} \right],\\
\delta \tilde E_g^{(\phi)}=\delta \left( E_g^{(\phi)} \dot \phi /Ha^3\right) &=& \frac{\h_g}{2} \left(\frac{\n+3}{\n+4}\right)  \left[ \ \left( \frac{\delta H^2}{H^2}  \right)'+ \frac{\delta H^2}{H^2}  +  \left( \frac{\delta \phi}{\phi}\right)'+(\n+3) \frac{\delta \phi}{\phi} \right], \\
\delta \tilde E_r^{(\phi)}=\delta \left( E_r^{(\phi)} \dot \phi /Ha^3\right) &=& \frac{\h_r}{2}\left[  \left( \frac{\delta H^2}{H^2}  \right)'-2 \frac{\delta H^2}{H^2}  - \left( \frac{\delta \phi}{\phi}\right)'-(\n+6) \frac{\delta \phi}{\phi} \right].
\ea
Note that the unperturbed Hamiltonian densities vary as $\h \propto a^\n$, which suggests that even a large  cosmological constant may be regarded as small perturbation, provided $\rho_\Lambda/a^\n \ll 1$. We now use (\ref{hampert}) to find $\delta H^2/H^2$, and substitute the result into (\ref{ephipert}). In each case we obtain 
\be
\left( \frac{\delta \phi}{\phi}\right)''+\left(\n+\frac{9}{2} \right)\left( \frac{\delta \phi}{\phi}\right)' =f(\n; c_1, c_2)\frac{ \rho_\Lambda}{(\nu a)^\n},
\ee
where
\be
f(\n; c_1, c_2)=\begin{cases} -\frac{2(2\n+27))(2\n+7)^3c_2^2}{(2\n+9)^2(4\n^2+28\n+9)c_1^3} & \textrm{Matter I} \\
-\frac{(2\n+9)^2(2\n^2+5\n-9)c_1}{16(2\n+7)(4\n^3+32\n^2+71\n+27)c_2^2}
& \textrm{Matter II}
\end{cases}
\ee
This is straightforwardly solved to give
\be
\frac{\delta \phi}{\phi}=A_1+ \frac{A_2}{a^{\n+\frac{9}{2}}}-\frac{2f}{9\n}  \frac{ \rho_\Lambda}{(\nu a)^\n},
\ee
where $A_1$ and $A_2$ are integration constants. It immediately follows that 
\be
\frac{\delta H^2}{H^2}=B_1+ \frac{B_2}{a^{\n+\frac{9}{2}}}+g  \frac{ \rho_\Lambda}{(\nu a)^\n},
\ee
where the constants $B_1, B_2$ and $g$ are related to $A_1, A_2$ and $f$ respectively. It is now clear that for $\hat n>0$,  $c_1, c_2 \neq 0$, the solutions (\ref{sol1}) and (\ref{sol2}) are late time attractors for vanishing curvature, even when we switch on $\rho_\Lambda$.  This view is reinforced by the plots shown in figures \ref{fig-matter} and \ref{fig-matterii}. In each case, the evolution mimics a matter dominated epoch for a long time even in the presence of a cosmological constant. When we also include curvature in the numerical simulation,  the solution ultimately  gives in to self tuning and asymptotes to a Milne universe.  

\begin{figure}[h] 
  \centering
  \includegraphics[width=3in]{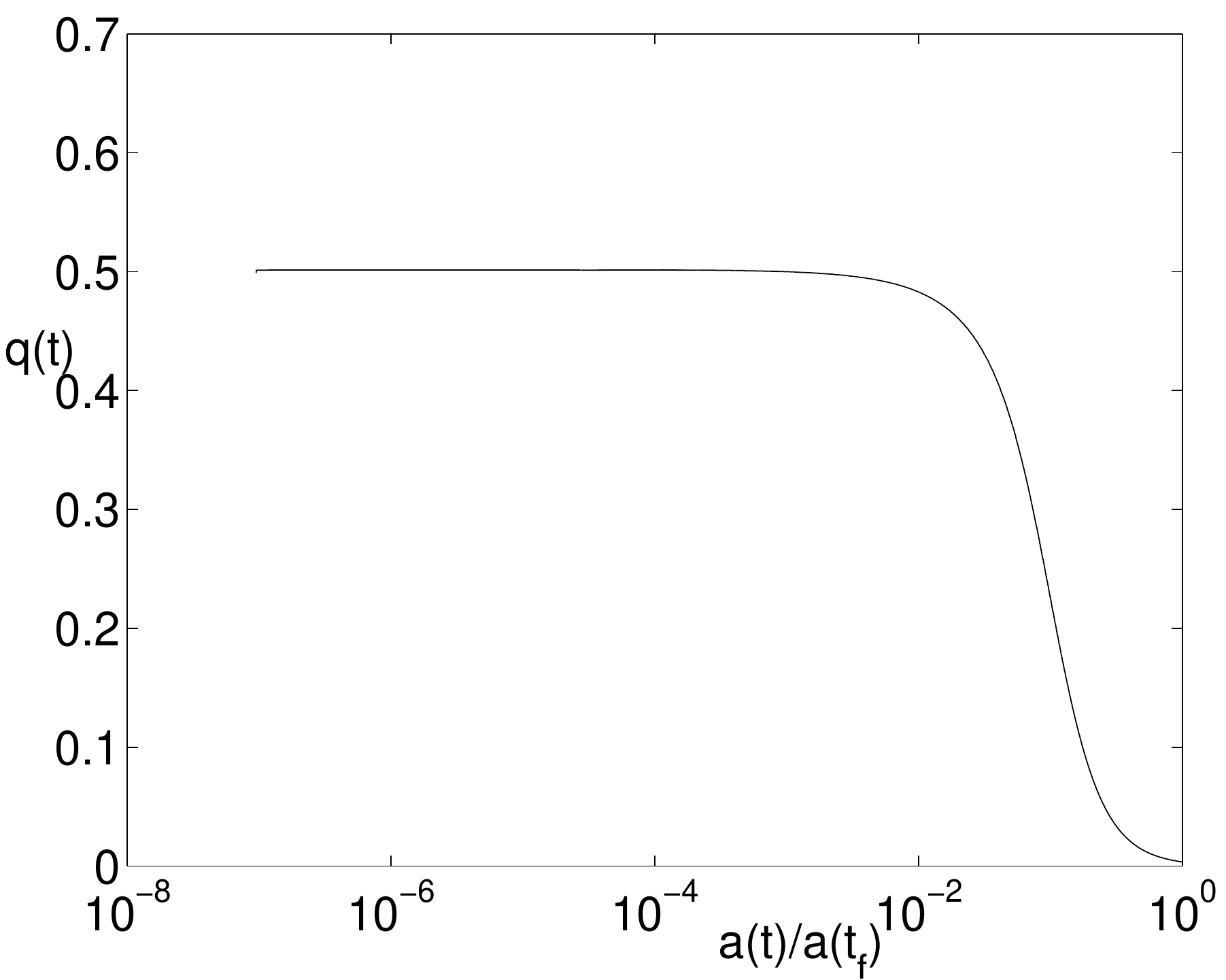} 
  \caption{Plot of the deceleration parameter for an initially Matter III solution, with $k=-0.1$, $\rho_\Lambda=1000$, $\phi_{initial}=0.1$, $c_1=1$.}
    \label{fig-matter3}
\end{figure}

The third class of potentials to mimic matter domination, which we dub ``Matter III", are given by 
\be
V_j (\phi)=-\frac{1}{2} c_1 \phi^{4}, \qquad  V_p(\phi)=V_g(\phi)=0, \qquad V_r(\phi)=\frac{1}{16} c_1 \phi^{6},
\ee
and follow from the "arbitrary" row of table \ref{tab-cosmologies}, giving a matter-like solution for arbitrary $\rho_\Lambda$ when $h=-3/2$ and $\sigma=0$. The explicit solution is
\be \label{sol3}
\phi=\nu a,  \qquad H^2=\sqrt{\frac{2 \rho_\Lambda}{81 c_1 }} (\nu a)^{-3},
\ee
where $\nu$ is an arbitrary constant.  The numerical solution with non-vanishing curvature is shown  in figure \ref{fig-matter3}. 

\begin{figure}[h] 
  \centering
  \includegraphics[width=3.5in]{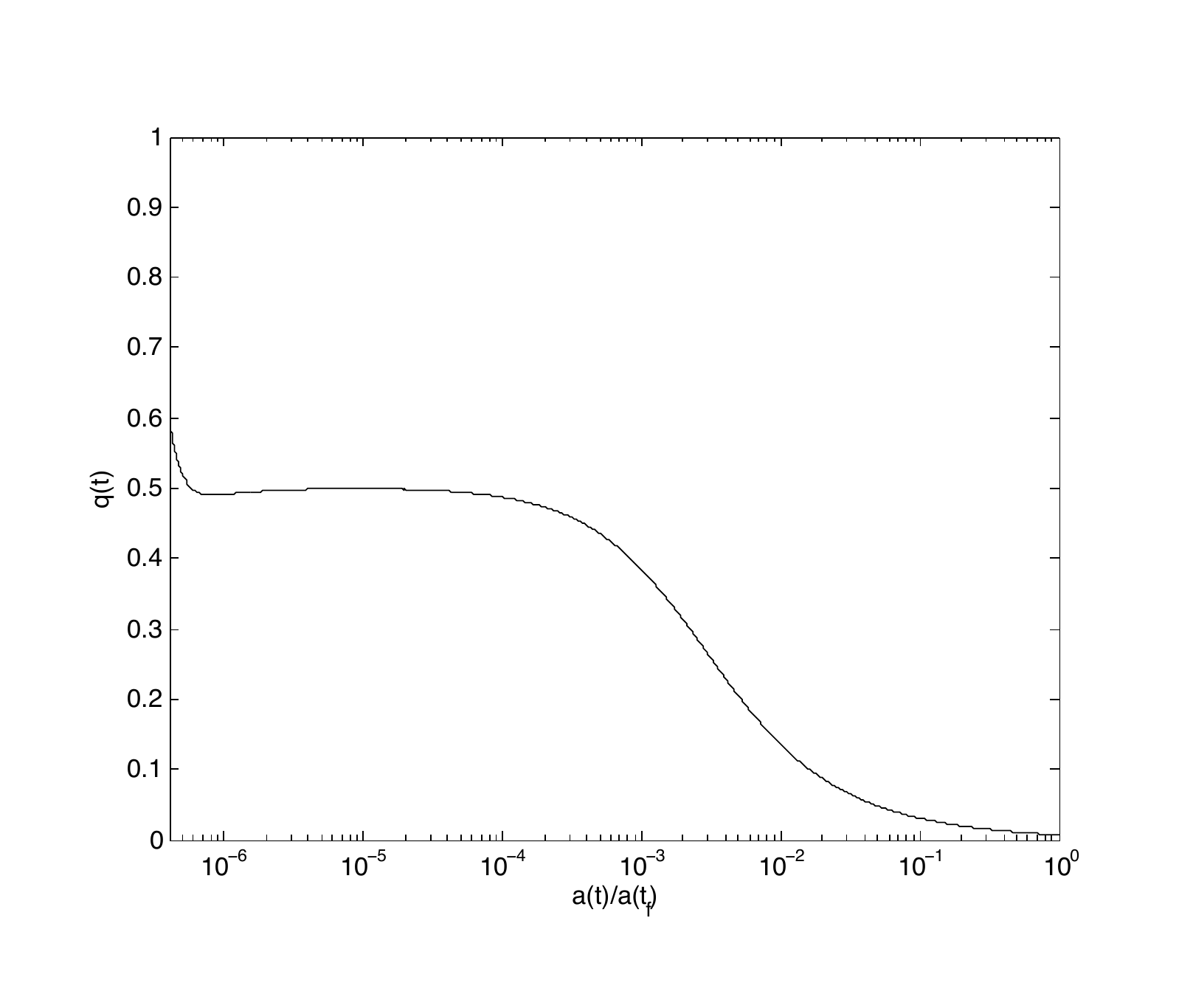} 
  \caption{Plot of the deceleration parameter for the Matter IV solution, with $k=-10^{-7}$, $\phi_{initial}=3.0$, $\rho_\Lambda=0.1$.}
    \label{fig:qMatter4}
\end{figure}

The fourth class of matter dominated solutions correspond to the case where we seek a single common root to the equations (\ref{eq:ham-n-sigma=0-2T}) and (\ref{eq:ham-n-sigma=0-2T}). An example, which we dub ``Matter IV'', is given by
\be
V_j (\phi)=-45\sqrt{2} \phi^{5}, \qquad  V_p(\phi)=-\frac{75067}{225}\phi^7, \qquad V_g(\phi)=-\phi^4, \qquad V_r(\phi)=\frac{143}{168}\sqrt{2}\phi^7
\ee
A plot of the resulting evolution is shown in figure \ref{fig:qMatter4}.

In summary, then, we have found four classes of potential that mimic a matter dominated cosmology, even when the source is dominated by a cosmological constant.  The \FF potentials have forced the scalar to screen the pressure component of the cosmological constant before it screens the energy density. This allows for an intermediate, and in some cases, pathology-free, period resembling a matter dominated cosmology, even in the absence of a pressureless source. At the latest times, the energy component of the cosmological constant is also screened and the solutions evolve to an asymptotically Milne universe, which is, of course, equivalent to a patch of  Minkowski space.
\subsubsection{Stability during matter epoch}\label{stability}
As promised, we will now consider cosmological perturbations about each class of matter solution.  Since we are interested in the phase prior to self-tuning, we shall restrict attention to vanishing curvature, and the solutions given by (\ref{sol1}), (\ref{sol2}) and (\ref{sol3}). Vacuum perturbations  about spatially flat cosmologies were studied in detail for Horndeski's theory in \cite{Kobayashi:2011nu}. Working  in unitary gauge, the scalar takes on its background value, $\phi=\phi(t)$, whereas the line element is given by
\be
ds^2=-N^2 dt^2+\gamma_{ij} (dx^i+N^i dt) (dx^j+N^j dt),
\ee
with 
\be
N=1+\hat{\alpha}, \qquad N_i=\del_i \hat{\beta}, \qquad \gamma_{ij}=a^2(t) e^{2\zeta} \left( \delta_{ij} +h_{ij} \right).
\ee
The quadratic action for tensor perturbations is found to be
\be
S_{tensor}=\frac{1}{8} \int dt d^3 x a^3 \left[ {\cal G}_T \dot h_{ij}^2 -\frac{{\cal F}_{T}}{a^2} (\vec \nabla h_{ij})^2 \right],
\ee
where ${\cal G}_T$ and ${\cal F}_T$ depend on the potentials evaluated on the background solution. Note that \cite{Kobayashi:2011nu} uses the DGSZ form of Horndeski's theory \cite{general}, so ${\cal G}_T$ and ${\cal F}_T$ are given in terms of DGSZ potentials.  The DGSZ potentials for the \FF are given in  Appendix C of \cite{Charmousis:2011ea}, so we can use this to extract the form of ${\cal G}_T$ and ${\cal F}_T$ for the cases we are interested in. We shall spare the reader the details, since they add little to this discussion.

Upon integrating out the lapse and the shift, the scalar perturbations yield the following effective action, 
\be
S_{scalar}=\int dt d^3 x a^3 \left[ {\cal G}_S \dot \zeta ^2 -\frac{{\cal F}_{S}}{a^2} (\vec \nabla \zeta)^2 \right].
\ee
Again, the coefficients ${\cal G}_S$ and ${\cal F}_S$ depend on the DGSZ potentials evaluated on the background solution. 

Now, to avoid a ghost instability, we require
\be
{\cal G}_T>0, \qquad {\cal G}_S>0,
\ee
and to avoid a gradient instability, we require 
\be
{\cal F}_T>0, \qquad {\cal F}_S>0.
\ee
It turns out that the Matter I, II and III potentials  do not give rise to well behaved cosmological perturbations about the relevant background solution (\ref{sol3}).  For example, for Matter III tensor perturbations are fine, with no ghost or gradient instability but the scalar perturbations exhibit a gradient instability
\be
{\cal G}_S >0,\qquad {\cal F}_S<0
\ee
For modes of wavelength $\lambda$ this instability manifests itself on timescales $t_{instability} \sim \lambda /|c_s| $, where $|c_s|=\sqrt{\left|\frac{{\cal F}_S}{{\cal G}_S}\right|} \sim\order(1)$. This is far too quick, and rules out the Matter III solution as part of a viable cosmology. 

In contrast, the Matter IV potentials do give rise to stable cosmological perturbations. About the exact solution with vanishing cosmological constant, we find that we have stable tensor fluctuations
\be
{\cal G}_T \sim \order(1)a^4 ,\qquad {\cal F}_T \sim \order(10)a^4
\ee
and stable scalar fluctuations
\be
{\cal G}_S \sim\order(10)a^4 ,\qquad {\cal F}_S \sim \order(10)a^4.
\ee
\section{Conclusions}\label{conc}
Obtaining a sensible cosmology in the presence of a large and changing contribution to the vacuum energy $\rho_\Lambda$ is one of major challenges facing the self tuning scenario that we have developed in \cite{Charmousis:2011bf,Charmousis:2011ea}. A standard cosmological evolution arising out of General Relativity with large $\rho_\Lambda$ would be totally unacceptable apart from perhaps in the early Universe where it would drive a period of accelerated expansion. There would be no way of exiting this period of inflation and obtaining a period of radiation and matter domination required for nucleosynthesis and structure formation. The goal of this paper is to demonstrate that for the \FF,  it is indeed possible to obtain  a sensible cosmological history even in the presence of a large $\rho_\Lambda$ contribution at all times. In other words, the self tuning of the cosmological constant can be accommodated in a sensible cosmological timeline. 

To show this we have developed a dynamical systems approach in which fixed point solutions corresponding to inflationary, radiation and matter dominated solutions are made manifest. Two key approaches are developed. In the first, we explicitly include the $\rho_\Lambda$ contribution and by demanding that all contributions in the Hamiltonian constraint remain independent of the Hubble parameter, we show that there exist a class of scaling solutions corresponding to the cosmologies we are looking for (recall there is no requirement here to actually include matter or radiation sources, the scalar field is doing the work for us). However, it turns out that this particular matter dominated solution (called ``Matter III"), whilst perfectly acceptable at the background level, actually contains a gradient instability when perturbed, an instability that would grow on too fast a timescale to be compatible with observations. This has led us to consider a second complementary approach. Rather than obtain background solutions in the presence of $\rho_\Lambda$ we set it to zero and look for consistent solutions that can also mimic  matter domination ($H^2 \propto a^{-3}$). Our requirement that the Hamiltonian constraint be independent of the Hubble parameter is now lifted and this allows for more freedom introducing an extra parameter (we call $n$ or equivalently, $\hat n $) in the background solutions.  We find  new classes of matter-like solutions (called ``Matter I" , `Matter II" and ``Matter IV") .

It would seem to go against the self-tuning spirit of the \FF that the ``Matter I" and ``Matter II" solutions only correspond to fixed points for vanishing cosmological constant. However, we have shown that they can still represent an excellent approximation even when a large $\rho_\Lambda$ is turned on, provided $\hat n>0$.  This is because the solution gets corrected by $\rho_\Lambda/a^{\hat n}$, a correction that decreases with time as the scale factor grows.   Using analytic methods,  we also showed  that for vanishing spatial curvature,  these solutions are cosmological attractors for $\hat n>0$. Once spatial curvature is reintroduced alongside the cosmological constant, we are forced to use numerical simulations which reproduce the expected behaviour: a long period of matter-domination, before asymptoting to the self-tuning Milne Universe. 

In contrast to the  Matter I, II, and III potentials, the Matter IV potentials have  corresponding solutions that are stable against cosmological perturbations.  This opens  up the possibility of a sensible matter dominated period of evolution, hence of structure formation in the {\it  Fab-Four} scenario.  Furthermore, these solutions are behaving in such a way that the scalar screens the pressure component of the cosmological constant before the energy density. At least for homogeneous and isotropic backgrounds, this suggests that the cosmological constant is being forced to behave like cold dark matter.  It is certainly tempting to ask whether such behaviour extends to inhomogeneous solutions, and recent results suggest that it may well be possible to have a \FF scenario satisfying current solar system constraints \cite{Rinaldi:2012vy}.

There is much that remains to be done. We have not yet obtained a full cosmology, but the fact that we have a class of background polynomial potentials that we know can provide the various cosmological epochs we want to reproduce offers us some direction.  Indeed we can speculate as to how we might sew together these interesting potentials to achieve the desired results. The point is that the scalar field is continually evolving, so we could arrange for the potential to correspond to different fluid behaviours for different ranges of $\phi$. For example, if we want radiation domination for $H^2>H^2_{eq}$ and  ``Matter I" like behaviour for $H^2<H^2_{eq}$, we might propose a Lagrangian of the form
\be
{\cal L}= \frac{m^4}{H_{eq}^2} \phi^{\hat n} R+ {\cal L}_{\textrm{``Matter I"}}\Big |_{c_1 \sim \order(1), c_2 \sim \order(m^{-2})},
\ee
Plugging in the ``Matter I' solution (\ref{sol1}), we see that
\be
\frac{m^4}{H_{eq}^2} \phi^{\hat n} R \sim m^4 a^{\hat n} \left(\frac{a_{eq}}{a} \right)^3, \qquad {\cal L}_{\textrm{``Matter I"}}\Big |_{c_1 \sim \order(1), c_2 \sim \order(m^{-2})} \sim  m^4 a^{\hat n},
\ee
where $a_{eq}=(m/H_{eq})^{2/3}$ is the value of the scale factor when $H=H_{eq}$. We see that the ``Matter I" terms dominate for $a>a_{eq}$, as desired for mater domination.  For $a<a_{eq}$ the $\phi^{\hat n}R$ term becomes important, and might be expected to dominate the dynamics, yielding an earlier  period of radiation domination. The same may be done to evade the graceful exit problem due to the inflationary solutions being attractors.

\begin{acknowledgments} 
We would like to thank Christos Charmousis for collaboration during the early stages of this work. EJC acknowledges  financial support from the Royal Society, STFC and Leverhulme Trust. AP  acknowledges  financial support from the Royal Society and PS from STFC.
\end{acknowledgments}

\end{document}